\documentstyle[amsfonts,eqsecnum,prd,aps]{revtex}

\begin{document}

\title{Rigid and gauge Noether symmetries for constrained %
systems}

\author{J.\ M.\ Pons
\footnote[1]{Electronic address: pons@ecm.ub.es}}
\address{Departament d'Estructura i Constituents de la Mat\`eria, 
Universitat de Barcelona,\\
and Institut de F\'\i sica d'Altes Energies,\\
Diagonal 647, E-08028 Barcelona, Catalonia, Spain\\
 and\\
   Center for Relativity, Physics Department, \\
   The University of Texas, Austin, Texas 78712-1081, USA}

\author{J.\ Antonio \ Garc\'{\i}a
\footnote[2]{Electronic address: garcia@nuclecu.unam.mx}}
\address{Instituto de Ciencias Nucleares\\
   Universidad Nacional Aut\'onoma de M\'exico\\
   Apartado Postal 70-543, 04510 \\
   M\'exico, D.F.}

\date{August 25, 1999}
\maketitle



\begin{abstract}
We develop the general theory of Noether symmetries for constrained 
systems, that is, systems that are described by singular Lagrangians. 
In our derivation, the Dirac bracket structure with respect to the 
primary constraints appears naturally and plays an important role in 
the characterization of the conserved quantities associated to these 
Noether symmetries. The issue
of projectability of these symmetries from tangent space to phase 
space is fully analyzed, and we give a geometrical interpretation of 
the projectability conditions in terms of a relation between the 
Noether conserved quantity in tangent space and the presymplectic 
form defined on it. We also examine the enlarged formalism that 
results from taking the Lagrange multipliers as new dynamical 
variables; we find the equation that characterizes the Noether 
symmetries in this formalism, and we also prove that the standard formulation
is a particular case of the enlarged one. The algebra of 
generators for Noether symmetries is discussed in both the 
Hamiltonian and Lagrangian formalisms. We find that a frequent source 
for the appearance of open
algebras is the fact that the
transformations of momenta in phase space and tangent space only 
coincide on shell. Our results apply with no distinction to rigid and 
gauge symmetries; for the latter case we give a general proof of 
existence of Noether
gauge symmetries for theories with first and second class constraints 
that do not exhibit tertiary constraints in the stabilization 
algorithm. Among some examples that illustrate our results, we study 
the Noether gauge symmetries of the Abelian Chern-Simons theory in 
$2n+1$ dimensions.  An interesting feature of this example is that its
primary constraints can only be identified after the determination of the
secondary constraint. The example is worked out retaining all the 
original set of variables. 

\end{abstract}

\pacs{11.10.Ef, 11.30.-j, 04.20.Fy  \hfill hep-th/9908151}
\twocolumn

\section{Introduction}

Variational principles and symmetries are two of the most fundamental 
organizing concepts in theoretical physics. One can add to this list 
the gauge principle. Noether symmetries, that is, continuous 
transformations that leave the action invariant up to boundary terms, 
constitute a link between the first two principles. In the case of a 
given gauge theory, some of the existing Noether transformations 
exhibit special features (the presence of Noether identities, 
arbitrary functions, etc.) that are associated with the redundancy in 
the description of degrees of freedom and the structure of 
constraints that characterize these theories. In a gauge theory there 
are two types of Noether symmetries, rigid and gauge. The first ones 
are physical symmetries with standard conserved quantities; the 
second, the gauge symmetries, are unphysical, and their associated 
conserved quantities vanish on-shell. We will deal simultaneously 
with both types of symmetries. 

Noether transformations provide for a wide class of symmetries of the 
equations of motion at the classical level. They also become, in the 
absence of anomalies, the symmetries of the quantum systems, and they 
are neatly displayed in the path integral formulation. But despite of 
the relevance of these symmetries for physical systems, many aspects 
of their characterization as well as the characterization of their 
conserved
quantities, either in the tangent space or in the phase space of some 
configuration space, have been insufficiently studied. There is at 
least one reason: It
is neither immediate nor trivial to extend the results that one can 
obtain for regular theories (theories with no gauge invariances) to 
theories described through singular Lagrangians (thus having room for 
gauge invariance). Our aim is to contribute to such a study. 

We will first focus in the first place (more general cases will be 
dealt with later) on infinitesimal Noether transformations of the 
type $\delta q (q,\, \dot q;\,t)$ in the tangent bundle $TQ$ of some 
configuration space $Q$, extended to include the independent variable 
(time) to $TQ \times R$, for some theory whose dynamics is described 
by a first order variational principle based on a time-independent 
Lagrangian $L(q,\dot q)$. Our transformations act only on the 
dependent variables $q$; this is general enough because any 
infinitesimal transformation which also acts on the independent 
variable $t$ (or the space-time variables in field theory) can be 
brought to this form.
The defining property of these transformations is that they leave the 
Lagrangian invariant up to a total time derivative: \begin{equation}
\delta L = \frac{d\,F}{d\,t}.
\label{dell}
\end{equation}
(In the case of field
theory the total time derivative is substituted by a total space-time 
divergence) We use the language most common to physics. In a more 
mathematically oriented language, see \cite{olver}, $\delta L$ would 
be understood as the action on $L$ of the prolongation of the vector 
field that generates the transformation. Except for an infinitesimal 
parameter, $\delta q$ is the characteristic of such a vector field. 

Equation (\ref{dell}) guarantees that $\delta q$ maps solutions of 
the equations of motion into solutions, because the equations of 
motion remain invariant. Associated with this Noether transformation 
there is always a conserved quantity,
$$
G_L = ({\partial L}/{\partial
\dot q}) \delta q - F.
$$
If the Lagrangian is non-singular, that is, if $\det |{\partial 
L}/{\partial \dot q \partial \dot q}| \not= 0$, then velocities are 
mapped one-to-one to canonical momenta, and the conserved quantity 
$G_L$ becomes in phase space the canonical generator, $G$, acting
through the Poisson bracket, of the transformation $\delta q$. 
However, the most common case in theoretical physics is the case when 
$L$ is singular. It is only under this circumstance that the 
phenomena of gauge freedom may occur. For a singular $L$, $G_L$ is 
still a projectable quantity \cite{kami82}, that is, it may be 
brought to the phase space as a function $G$. This function $G$ is 
now uniquely defined only up to
the addition of linear combinations of the primary constraints. In 
contrast with the non-gauge case, the fact of being a canonical 
conserved quantity does not guarantee that $G$, or any of its 
equivalent functions whose pull-back to tangent space is $G_L$, 
generates $\delta q$ or even a Noether transformation. It is not even 
guaranteed that $\delta q$ is a transformation projectable to phase 
space.

Our study will clarify in what sense, and to what extent, we may 
consider $G$ as the generator of the original transformation $\delta 
q$. We give a general characterization of the functions $G$ in phase 
space that are associated with a Lagrangian Noether transformation in 
the case of gauge theories, and we show the general construction, 
made out of $G$, of these transformations. We prove that the 
projectability to phase space of $\delta q$ is related to a geometric 
requirement that $G_L$ must satisfy. When this requirement is met, we 
show that $\delta q$ is canonically generated. 

Our procedure shows that there is a natural generalization to a 
framework, the enlarged formalism, where the dynamics is defined 
through the canonical Lagrangian. In this enlarged formalism the 
Lagrange multipliers associated to the primary Hamiltonian 
constraints become new independent variables. Then we can define 
generalized Noether transformations depending on the Lagrange 
multipliers and their time derivatives at any finite order. We make 
contact here with the formulation in \cite{htz,ht}. It turns out 
that the results in \cite{htz}
can be understood as an application of our formulation in order to 
generate Noether gauge transformations for systems with only first 
class constraints in a systematic way.

Our regularity assumptions are standard: We consider that the Hessian 
matrix of the Lagrangian with respect to the velocities has constant 
rank. We also assume that the primary constraints may be split into 
first and second class on the primary constraint surface. Some of our 
results, particularly in section 4, will not depend on this second 
assumption. 

The paper is organized as follows:
In section 2 we introduce some known results on the dynamics of 
Lagrangian and Hamiltonian systems that will be needed in the 
following sections. In section 3 we develop the general theory of 
Noether symmetries for gauge systems. In particular we characterize 
the conserved quantities in phase space associated with these 
symmetries and we give the method
to retrieve the original Noether transformation out of its conserved 
quantity. It is remarkable that the Dirac bracket structure plays a 
natural role in this context. We also relate the projectability 
conditions for the Noether transformation to a property of the 
Lagrangian Noether conserved quantity $G_L$ that relates to
the presymplectic form in tangent space. We also see that a 
projectable Noether transformation always becomes a canonical 
transformation in phase space.

In section 4 we introduce
the enlarged formalism with the canonical Lagrangian. The definition 
of Noether transformations is then generalized by allowing $\delta q$ 
to depend on the Lagrange multipliers \cite{htz}, and we prove the 
equivalence with the former formulation when this dependence does not 
contain derivatives of the Lagrange multipliers. In section 5 some 
properties of the algebras of transformations and generators are 
exhibited. In particular we show that the closure under Poisson 
bracket of the algebra of generators in phase space does not 
necessarily guarantee the closure of the algebra of the infinitesimal 
transformations in configuration space. This fact is relevant for the 
application of BRST methods \cite{BV}. In section 6 we distinguish 
between rigid and gauge Noether symmetries, and extend some theorems 
concerning the existence of canonical gauge Noether transformations 
to systems with first and second class constraints with only one step 
in the stabilization algorithm. Finally, we present some examples to 
illustrate our results and a short appendix devoted to the concept of 
auxiliary variables that will be used in section 4. Of particular 
interest is example 4, devoted to the analysis of the Noether gauge 
symmetries of the Abelian Chern-Simons theory in $2n+1$ dimensions. 
The example is worked out retaining all the original set of 
variables. It is interesting to remark that this example violates one 
of our regularity assumptions, but we show that it can be still 
accommodated within our formalism.

Our results are local. They apply to systems with a finite number of 
degrees of freedom as well as to field theories. We use, for 
simplicity, the language of mechanics. DeWitt's condensed index 
notation \cite{dewitt} translate our results directly to field theory 
as long as the boundary conditions allow for the elimination of 
surface terms. 

We rely on previous work, particularly \cite{equiv,b-g-g-p89,evol} 
and references therein, and so we must
first summarize some of the results of these papers before proceeding 
to subsequent developments.


\section{Elements of Lagrangian and Hamiltonian dynamics for gauge 
theories}

The Euler-Lagrange functional derivative of a first order Lagrangian 
$L(q,\, \dot q)$, is
$$
[L]_i := \alpha_i - W_{is}\ddot q^s,
$$
where
$$
W_{ij} := {\partial^2L\over\partial\dot q^i\partial\dot q^j} \quad 
{\rm{}and} \quad \alpha_i :=
- {\partial^2L\over\partial\dot q^i\partial q^s}\dot q^s + {\partial 
L\over\partial q^i} .
$$
We consider the general case where the Hessian ${\bf{}W}=(W_{ij})$ 
may be a singular matrix \cite{dirac}. We assume that its rank is 
constant in the region of tangent space of our interest. If 
${\bf{}W}$ is singular, there exists a kernel for the pullback ${\cal 
F}\!L^*$ of the
Legendre map ${\cal F}\!L$ from configuration-velocity space $TQ$ 
(the tangent bundle $TQ$ of the configuration space $Q$) to phase 
space $T^*Q$ (the cotangent bundle). This kernel is spanned by the 
vector fields
\begin{equation}
{\bf\Gamma}_\mu
= \gamma^i_\mu {\partial\over\partial\dot q^i} , \label{GAMMA}
\end{equation}
where $\gamma^i_\mu$ span a basis for the null vectors of $W_{ij}$. 
\footnote{Notice that if the phase space is enlarged by introducing 
the Lagrange multipliers as new dynamical variables, the accordingly 
enlarged Legendre map will now become invertible. Therefore we expect 
that the problems
related with projectability can be overcome in this enlarged 
formalism. We will consider this issue in section 4.} A function 
$g(q,\,\dot q)$ is projectable to phase space if and only if $$
{\bf \Gamma}_\mu g = 0.
$$
Notice that the Lagrangian equations of motion $[L]_i =0$ imply the 
primary Lagrangian constraints
$$
\chi_{\mu} :=(\alpha_i \gamma^i_\mu) =0. $$

The time evolution for a gauge theory is not unique until the gauge 
freedom has been removed, for example, by way of some gauge fixing. 
This is reflected in the ambiguities present in the Lagrangian 
time-evolution differential operator \cite{equiv}: \begin{equation}
{\bf X_L} := {\partial\over\partial t}
+ \dot q^{s}{\partial\over\partial q^{s}} +a^{s}(q,\dot 
q){\partial\over\partial \dot q^{s}} +\eta^{\mu}{\bf\Gamma}_{\mu} =: 
{\bf X}_{0} +\eta^{\mu}{\bf\Gamma}_{\mu} , \label{EVOLOP} 
\end{equation}
where $a^{s}$ are functions which are determined by the formalism, 
and $\eta^{\mu}$ are arbitrary functions. These arbitrary functions 
express the gauge freedom of the time-evolution operator. Notice that 
projectable quantities have a well defined unambiguous dynamics. The 
tangency of $\bf X_L$
to the primary Lagrangian constraint surface, defined by $\chi_{\mu} 
=0$, may lead to new constraints and to the determination of some of 
the functions $\eta^\mu$. At this point, new tangency requirements 
may occur \cite{dirac,SM}.

It is not necessary to use the Hamiltonian technique to find the 
${\bf\Gamma}_\mu$, but it does
facilitate the calculation:
\begin{equation} \gamma^i_\mu
= {\cal F}\!L^*\left( \partial\phi_\mu\over\partial p_i \right) , 
\label{gam}
\end{equation}
where the $\phi_\mu$ are the Hamiltonian primary constraints. They 
satisfy by definition ${\cal F}\!L^* \phi_\mu = 0$.

These constraints $\phi_\mu$ span a basis for the ideal of functions 
in $T^*Q$ that vanish on the image of the Legendre map ${\cal F}\!L$. 
We take as an assumption that these constraints may be split into 
first
class, $\phi_{\mu_0}$, and second class, $\phi_{\mu_1}$, 
\footnote{Exceptions to this assumption can be still accomodated 
within our formalism, as example 4 of section 7 shows.}
satisfying
\begin{equation}
\{ \phi_{\mu_0},\,\phi_{\mu} \} = pc, \quad \det |\{ 
\phi_{\mu_1},\,\phi_{\nu_1} \}| \not= 0, \label{fc-sc} \end{equation}
where $\{-,\,- \}$ is the Poisson Bracket structure and $pc$ stands 
for a generic linear combination of the primary constraints.

The Lagrangian energy,
$E_L(q, \dot q) :=
\dot q^i (\partial L / \partial \dot q^i) - L$, is a function 
projectable to phase space. The canonical Hamiltonian $H_c(q,p)$, 
which is only uniquely defined up to primary constraints, is defined 
such that its pullback to tangent space is the Lagrangian energy:
$$
{\cal F}\!L^* H_c = E_L.
$$

To connect the Lagrangian and Hamiltonian dynamics it is convenient 
to write down the two following identities \cite{equiv}: \begin{equation} \dot 
q^i = {\cal F}\!L^* ({\partial H_c\over\partial p_i}) + v^\mu(q, \dot 
q) {\cal F}\!L^*
({\partial \phi_\mu\over\partial p_i}),
\label{id1}
\end{equation}
and
\begin{equation}
{\partial L\over\partial q^i}
= - {\cal F}\!L^* ({\partial H_c\over\partial q^i}) - v^\mu(q, \dot 
q) {\cal F}\!L^*
({\partial \phi_\mu\over\partial q^i});
\label{id2}
\end{equation}
where the functions $v^\mu$ are
determined so as to render the first relation an identity. Notice the 
important relation
\begin{equation}
{\bf\Gamma}_\mu v^\nu = \delta_\mu^\nu,
\label{delta}
\end{equation}
which stems from applying ${\bf\Gamma}_\mu$ to the first identity and 
taking into account that
\begin{equation} {\bf\Gamma}_\mu \circ{\cal F}\!L^* = 0, \label{zero}
\end{equation}
where $\circ$ denotes the composition operation. 

The Hamiltonian time evolution vector field is given by \begin{equation} {\bf 
X_H}:={\partial \over \partial t} + \{-,\,H_c \} + \lambda^\mu \{-,\, 
\phi_\mu \}, \label{xh} \end{equation}
where $\lambda^{\mu}$ are arbitrary functions of time. However, they 
are determined as non-projectable functions in tangent space: They 
are the functions $v^\mu(q, \dot q)$ implicitly defined by equations 
(\ref{id1}). These variables $\lambda^{\mu}$ are Lagrange multipliers.

The requirement of tangency
of $\bf X_H$ to the primary Hamiltonian constraint surface, defined 
by $\phi_\mu$, may
lead to new constraints and to the determination of some of the 
functions $\lambda^\mu$. This is the setting of the stabilization 
algorithm in phase space, that runs parallel (see \cite{equiv}) to 
the corresponding algorithm in tangent space. Details on the 
relationship between the functions $\eta^\mu$ in $TQ$ and the 
functions $\lambda^\mu$ in $T^*Q$ are given in \cite{pons88}. 

With the identities (\ref{id1}) and (\ref{id2}), we can relate the 
time evolution in the Hamiltonian and the Lagrangian formalisms for 
any function $f(q,\,p;t)$. In \cite{equiv} an evolution operator $K$ 
is defined that gives the time evolution of a function $f$ in 
$T^*Q\times R$ as a function in $TQ\times R$, \begin{equation}
K f := {\cal F}\!L^*{\partial f \over \partial t} + \dot q {\cal 
F}\!L^*{\partial f\over\partial q} + {\partial L \over \partial q} 
{\cal F}\!L^*{\partial f\over\partial p}. \label{k-rel} \end{equation}
This operator $K$ is fully studied in \cite{evol}. Notice the 
immediate result
\begin{equation}
K \phi_\mu = \chi_\mu,
\label{imm}
\end{equation}
which is deduced by recalling that ${\cal F}\!L^* \phi_\mu = 0$. 
Using the identities (\ref{id1}) and (\ref{id2}), we can get a new 
expression for $K f$,
\begin{equation}
K f = {\cal F}\!L^*{\partial f \over \partial t} + {\cal F}\!L^* 
\{f,\,H_c \} + v^\mu {\cal F}\!L^*\{f,\, \phi_\mu \}. \\ \end{equation}
With this new expression, application of the vector fields 
$\Gamma_\mu$
to (\ref{imm}) gives the following result \cite{equiv}: \begin{equation}
{\bf \Gamma}_\mu \chi_\nu =
{\cal F}\!L^* \{ \phi_{\nu},\,\phi_{\mu} \}. \label{gam-fl} \end{equation}

Now we are ready to relate the primary Lagrangian constraints to the 
secondary Hamiltonian constraints and the canonical determination of 
some arbitrary functions of the Hamiltonian dynamics. To this end, 
let us first apply (\ref{k-rel}) to $f = \phi_{\mu_0}$,
\begin{eqnarray}
\nonumber
\chi_{\mu_0} = K \phi_{\mu_0}
&=&{\cal F}\!L^* \{\phi_{\mu_0},\,H_c \}
+ v^\mu {\cal F}\!L^*\{\phi_{\mu_0},\, \phi_\mu \}\\
\nonumber
& =& {\cal F}\!L^* 
\{\phi_{\mu_0},\,H_c \},
\end{eqnarray}
where we have used (\ref{fc-sc}).
Then, if we define the secondary Hamiltonian constraints as \begin{equation}
\phi^1_{\mu_0} := \{\phi_{\mu_0},\,H \}, \label{second} \end{equation}
their pullback to tangent space gives a
subset of the primary Lagrangian constraints \footnote{Notice that 
our notation for the secondary Hamiltonian constraints may give some 
redundancy, for there is no guarantee that every $\phi^1_{\mu_0}$ 
defines a new independent constraint. This potential redundancy 
creates no problem within our formulation.}, \begin{equation}
\chi_{\mu_0}={\cal F}\!L^* \phi^1_{\mu_0}. \label{proj} \end{equation}

Next, applying
(\ref{k-rel}) to $f = \phi_{\mu_1}$, we find 
\begin{eqnarray}
\nonumber
\chi_{\mu_1} =
K \phi_{\mu_1} &=& {\cal F}\!L^* \{\phi_{\mu_1},\,H \} + v^\nu {\cal 
F}\!L^*\{\phi_{\mu_1},\, \phi_\nu \} \\
&=& {\cal F}\!L^* 
\{\phi_{\mu_1},\,H \}
+ v^{\nu_1} {\cal F}\!L^*\{\phi_{\mu_1},\, \phi_{\nu_1} \}. 
\label{nonpr}
\end{eqnarray}
The stabilization of the second class constraints fixes some 
arbitrariness in the Hamiltonian dynamics ${\bf X_H}$. The arbitrary 
functions $\lambda^{\nu_1}$ become determined as canonical functions 
$\lambda_c^{\nu_1}$ through \begin{equation}
0 =
\{\phi_{\mu_1},\,H_c \}
+ \lambda_c^{\nu_1} \{\phi_{\mu_1},\, \phi_{\nu_1} \}. \label{determ} 
\end{equation}
Then, we can put together (\ref{nonpr}) and (\ref{determ}) to get: 
\begin{equation}
\chi_{\mu_1} =
(v^{\nu_1} - {\cal F}\!L^* \lambda_c^{\nu_1}) {\cal 
F}\!L^*\{\phi_{\mu_1},\, \phi_{\nu_1} \}. \label{nonproj} \end{equation}

Notice that the first class and second class classification of the 
primary constraints has an effect in the tangent space formalism: it 
classifies the primary Lagrangian constraints into two sets according 
to their projectability to phase space.
Some of the Lagrangian constraints, $\chi_{\mu_0}$, are projectable, 
as shown by (\ref{proj}),
whereas the rest, the constraints $\chi_{\mu_1}$, are 
non-projectable, as we can verify by applying (\ref{delta}) and 
(\ref{zero}) to (\ref{nonproj}).
Notice also that the constraints of the type $\chi_{\mu_0}$, or some 
of
them, may vanish identically, whereas the constraints of the type 
$\chi_{\mu_1}$ are all independent.

This finishes the
summary of results that will be used in the next sections. 


\section{Noether symmetries}

\subsection{General theory of Noether symmetries in tangent space and 
in phase space}

We start with an infinitesimal
Noether Lagrangian symmetry on $TQ\times R$, that is, an 
infinitesimal $\delta q(q,\dot q;t)$ such that $$\delta L = 
\frac{d\,F}{d\,t},
$$
with
$$
\frac{d}{d\,t} = \frac{\partial}{\partial\,t} + {\dot 
q}\frac{\partial}{\partial\,q} +
{\ddot q}\frac{\partial}{\partial\, {\dot q}} + ... \ , $$
and we will investigate the conversion of this symmetry to the 
Hamiltonian formalism. Defining
\begin{equation}
G_L = (\frac{\partial L}{\partial \dot q^i}) \delta q^i - F, 
\end{equation}
we can write
\begin{equation} [L]_i\delta q^i + \frac{dG_L}{dt} = 0. \label{noet}
\end{equation}

Notice that the highest derivative in (\ref{noet}), $\ddot q$, 
appears linearly. Then, since (\ref{noet}) is identically satisfied 
for a Noether symmetry $\delta q$, the coefficient of $\ddot q^i$ 
must vanish:\footnote{
Notice that if $W_{is}$ is
invertible we can deduce from (\ref{w-g}) the form of the associated 
Noether transformation,
$$
\delta q^s=W^{is}\frac{\partial G_L}{\partial{\dot q}^i} $$
where $W^{is}$ denotes the inverse to $W_{is}$.} \begin{equation}
W_{is} \delta q^s - {\partial G_L \over\partial\dot q^i} = 0 . 
\label{w-g}
\end{equation}
We contract (\ref{w-g}) with a null vector $\gamma^i_\mu$ to find 
that \begin{equation} {\bf\Gamma}_\mu G_L = 0.
\label{gammg}
\end{equation}
It follows that $G_L$ is projectable to a function $G$ in~$T^*Q$; 
that is, it is the pullback of a function (not necessarily unique) in 
$T^*Q$:
$$
G_L = {\cal F}\!L^*(G) .
$$

This important property, valid for any conserved quantity associated 
with an infinitesimal Noether symmetry of the type considered here, 
was first pointed out in
\cite{kami82}. Observe that $G$ is determined up to the addition of 
linear combinations of the primary constraints. Substitution of this 
result in (\ref{w-g}) gives
$$
W_{is} \left( \delta q^s - {\cal
F}\!L^*
\left({\partial G\over\partial p_s}\right) \right) = 0 , $$
and so the parentheses enclose a null vector of {\bf W}: 
\begin{equation}
\delta q^i = {\cal F}\!L^*
\left({\partial G \over\partial p_i}\right) - \sum_\mu r^\mu 
\gamma^i_\mu ,
\label{dq}
\end{equation}
for some $r^\mu( q, \dot q; t)$.

Our aim is to get a complete characterization of the canonical 
generator $G$ in phase space. We start by defining \begin{equation}
\hat p_i = {\partial L\over\partial\dot q^i}. \label{hatp} \end{equation}
After subtraction from (\ref{noet}) of the piece containing $\ddot 
q^i$, we obtain
\begin{eqnarray}
\nonumber
\left({\partial L\over\partial q^i}
- \dot q^s {\partial \hat p_i\over\partial q^s} \right) \left( {\cal 
F}\!L^*({\partial G\over\partial p_i})
- \sum_\mu r^\mu \gamma^i_\mu \right)\\
+ \dot q^i {\partial \over\partial q^{i}} {\cal F}\!L^* (G)
+ {\cal F}\!L^* ({\partial G\over\partial t}) =0, 
\end{eqnarray}
which simplifies to
\begin{equation}
{\partial L\over\partial q^i} {\cal F}\!L^*( {\partial G\over\partial 
p_i} )
+ \dot q^i {\cal F}\!L^*({\partial G\over\partial q^i}) + {\cal 
F}\!L^*
({\partial G\over\partial t}) =r^\mu \chi_\mu. \label{noet-wg}
\end{equation}
Substitution of the two identities
(\ref{proj}) and (\ref{nonproj}) into
(\ref{noet-wg}) yields (see \cite{xgp92}) \begin{equation}
{\cal F}\!L^* ({\partial G\over\partial t}) + {\cal F}\!L^*\{G,H_c\} 
+ v^\mu {\cal F}\!L^*\{G,\phi_\mu \}
=r^\mu \chi_\mu.
\label{gra-po}
\end{equation}

Notice that (\ref{gra-po}) is invariant under \begin{equation}
r^\mu \rightarrow r^\mu +
s^\mu, \ {\rm where} \ \
s^\mu = A^{\mu\nu} \chi_\nu + s_0^\mu, \label{indet} \end{equation}
with $A$ an arbitrary antisymmetric matrix with functions in 
$TQ\times R$ as components, and with $s_0^\mu$ any function in 
$TQ\times R$ not vanishing at the Lagrangian primary constraint 
surface and such that
$s_0^\mu\chi_\nu$ is identically zero. Such $s_0^\mu$ only exist 
\cite{gra-po 94} for theories whose projectable primary Lagrangian 
constraints are not all independent. 

These indeterminacies for $r^\mu$ in (\ref{gra-po}) entail an 
irrelevant (trivial) change in the $\delta q^{i}$ of (\ref{dq}): \begin{equation}
\delta q^i \rightarrow \delta_1 q^i = \delta q^i - a^{ij}[L]_j - 
s_0^\mu\gamma_\mu^i,
\label{arbit}
\end{equation}
where $a^{ij} =\gamma_\mu^i A^{\mu\nu} \gamma_\nu^j$. This result is 
already obvious in (\ref{noet}) because $[L]_i \delta_1 q^i = [L]_i 
\delta q^i$.

Now we will obtain a purely Hamiltonian characterization of $G$. In 
doing so the relevance of the Dirac bracket structure will be 
emphasized.
Let us introduce $v^{\nu_1}$ of (\ref{nonproj}) into (\ref{gra-po}). 
We get
\begin{eqnarray}
{\cal F}\!L^* ({\partial G\over\partial t}) + {\cal 
F}\!L^*\{G,H_c\} + v^{\mu_0} {\cal F}\!L^*\{G,\phi_{\mu_0} 
\}\nonumber \\ + \left( {\cal
F}\!L^* \lambda_c^{\nu_1} + M^{{\nu}_1{\mu}_1} \chi_{\mu_1} \right) 
{\cal F}\!L^*\{G,\phi_{\nu_1} \} = r^\mu \chi_\mu,
\end{eqnarray}
where ${\bf M} = (M^{{\nu}_1{\mu}_1})$ is defined as the matrix 
inverse of the Poisson
bracket matrix of the primary second class constraints. Define $$
r'^{\mu_1} := r^{\mu_1} - {\cal F}\!L^*\{G,\phi_{\nu_1} 
\}M^{{\nu}_1{\mu}_1}
$$
and
$$
r'^{\mu_0} := r^{\mu_0}.
$$
We get
\begin{eqnarray}
\nonumber
{\cal F}\!L^* ({\partial G\over\partial t}) +{\cal F}\!L^*\{G,H_c\} + 
v^{\mu_0} {\cal F}\!L^*\{G,\phi_{\mu_0} \}  \\ + {\cal F}\!L^* 
\lambda_c^{\mu_1}
{\cal F}\!L^*\{G,\phi_{\mu_1} \}
= r'^\mu \chi_\mu.
\label{gra-po2}
\end{eqnarray}
However, $\lambda_c^{\mu_1}$ is determined from (\ref{determ}) as $$
\lambda_c^{\mu_1} = - M^{{\mu}_1{\nu}_1} \{\phi_{\nu_1},\,H_c \} . $$
Introduction of this result into (\ref{gra-po2}) produces the natural 
appearance of the Dirac bracket \cite{dirac}: \begin{equation}
{\cal F}\!L^* ({\partial G\over\partial t}) + {\cal 
F}\!L^*\{G,H_c\}^* + v^{\mu_0} {\cal F}\!L^*\{G,\phi_{\mu_0} \} = 
r'^\mu \chi_\mu, \label{gra-po3}
\end{equation}
where the Dirac bracket is defined, at this stage of the 
stabilization algorithm, by
$$
\{A,\,B\}^* := \{A,\,B\} - \{A,\,\phi_{\mu_1}\}M^{{\mu_1}{\nu_1}} 
\{\phi_{\nu_1},\,B\}.
$$

Now apply ${\bf \Gamma}_{\nu_0}$ to (\ref{gra-po3}). Using 
(\ref{delta}) and (\ref{zero}) we arrive at \begin{equation}
{\cal F}\!L^*\{G,\phi_{\mu_0} \} =
({\bf \Gamma}_{\mu_0} r'^\nu) \chi_\nu \label{part} \end{equation}
where we have also used (\ref{gam-fl}) and (\ref{fc-sc}) to get ${\bf 
\Gamma}_{\mu_0} \chi_\nu = 0.$ But since the left side of 
(\ref{part}) is a projectable function, so must be the
right. However, only the constraints $\chi_{\nu_0}$ are projectable. 
Indeed, according to (\ref{proj}), they are the pullback of the 
secondary Hamiltonian constraints. In conclusion, we can write that
\begin{equation}
\{G,\phi_{\mu_0} \} = sc + pc,
\label{cond1}
\end{equation}
where $sc$ ($pc$) stands for a linear combination of secondary 
(primary) Hamiltonian constraints.
Introduce this result in (\ref{gra-po3}); the same reasoning yields
\begin{equation}
{\partial G\over\partial t} + \{G,H_c\}^* = sc + pc. \label{cond2}
\end{equation}
Notice that (\ref{cond1}) can be equivalently written as \begin{equation}
\{G,\phi_{\mu_0} \}^* = sc + pc.
\end{equation}

Let us remark that the secondary constraints in the right sides of 
(\ref{cond1}) and (\ref{cond2}) are the stabilization of the primary 
first class constraints, as defined in (\ref{second}). Therefore we 
are not only saying that the left sides of (\ref{cond1}) and 
(\ref{cond2}) must vanish on the
surface defined by the primary plus secondary constraints. We are 
saying something more restrictive, because some of the secondary 
constraints, as obtained through the stabilization of the primary 
first class constraints according to (\ref{second}), may be 
ineffective (that is, such that their gradient also vanishes in the 
constraint surface). In such a case, the left sides of (\ref{cond1}) 
and (\ref{cond2}) must reflect this fact. Particular examples of this 
behavior can be found in \cite{ineffective}. 

We have arrived
at the following result: {\it The necessary and sufficient condition 
for a function $G \in T^*Q\times R$ to be a Noether canonical 
conserved quantity, that is, such that its pullback to $TQ \times R$ 
satisfies equation (\ref{noet}) for some $\delta q$, is that $G$ 
satisfies equations (\ref{cond1}) and (\ref{cond2}).}

This result generalizes to systems with
gauge freedom the standard definition of a Noether conserved quantity 
in phase space.


\subsection{Getting $\delta q$ from $G$} 

Equations (\ref{cond1}) and (\ref{cond2}) express the most general 
condition for a canonical conserved quantity to be Noether, that is, 
such that $G_L = {\cal F}\!L^* G$ satisfies (\ref{noet}) for some 
$\delta q^{i}(q, {\dot q}; t)$. In fact, if $G$ satisfies 
(\ref{cond1}) and (\ref{cond2}), then $\delta q^{i}$ can be obtained, 
except, of course, for the arbitrariness described in (\ref{arbit}), 
from $G$ as follows. Let us first rewrite (\ref{cond1}) and 
(\ref{cond2}) using a notation for the coefficients in the secondary 
constraints: \begin{equation}
{\partial G\over\partial t} + \{G,H_c\}^* = A^{\mu_0} \phi^1_{\mu_0} 
+ pc,
\label{cond2i}
\end{equation}
\begin{equation}
\{G,\phi_{\mu_0} \} = B^{\nu_0}_{\mu_0}\phi^1_{\nu_0} + pc. 
\label{cond1i}
\end{equation}
Notice that only the coefficients $B^{\nu_0}_{\mu_0}$ are invariant 
under the changes of $G$ allowed by
the addition of arbitrary linear combinations of the primary 
constraints \footnote{In the case of gauge symmetries, a redefinition 
of the arbitrary functions may help to achieve projectability, see 
section 6.2.}. Comparing (\ref{cond2i}) and (\ref{cond1i}) with 
(\ref{gra-po3}) and (\ref{part}), we identify a set of solutions for 
$r'^\mu$, $$
r'^{\mu_0} = {\cal F}\!L^* A^{\mu_0} + v^{\nu_0} {\cal 
F}\!L^*B^{\mu_0}_{\nu_0},
$$
$$
r'^{\mu_1} = 0,
$$
that give
\begin{equation}
r^{\mu_0} = {\cal F}\!L^* A^{\mu_0} + v^{\nu_0} {\cal 
F}\!L^*B^{\mu_0}_{\nu_0}
\label{erra0}
\end{equation}
and
\begin{equation}
r^{\mu_1} = {\cal F}\!L^*\{G,\phi_{\nu_1} \}M^{{\nu}_1{\mu}_1}. 
\label{erra1}
\end{equation}
The general solution for the $r^{\mu}$
may be obtained using (\ref{indet}).

Using (\ref{erra0}) and (\ref{erra1}) in (\ref{dq}) we find the 
result we were looking for,
\begin{equation}
\delta q^{i} = {\cal F}\!L^*\{q^{i},\,G \}^* - ({\cal F}\!L^* 
A^{\nu_0} + v^{\mu_0} {\cal F}\!L^* B^{\nu_0}_{\mu_0}) 
\gamma^{i}_{\nu_0}. \label{delq}
\end{equation}
Notice again the natural appearance of the Dirac Bracket. 

Up to now, $G$ is
any function in $T^*Q \times R$ whose pullback to $TQ\times R$ is 
$G_L$. We have obtained therefore the following results: First, we 
have in (\ref{cond1}) and (\ref{cond2}) the conditions for a function 
$G$ in phase space to be associated with a Lagrangian Noether 
transformation. Next, this transformation is entirely recovered 
through (\ref{delq}), up to the addition of trivial pieces of the 
type described in (\ref{arbit}). 

Notice that in general there are
obstructions that prevent $\delta q^{i}$ from being canonically 
generated. As we will see now, the
quantities $A^{\mu_0}$ are readily absorbed through a redefinition of 
$G$. Instead, the
quantities $B^{\nu_0}_{\mu_0}$ are the true obstructions to 
projectability: they cannot be absorbed in $G$ by the changes allowed 
because $G$ is only determined up to primary constraints. We define 
\begin{equation}
G' = G - A^{\mu_0} \phi_{\mu_0}
\label{gprime}
\end{equation}
and then
\begin{equation}
G'^* = G' - \{G',\,\phi_{\mu_1}\}M^{{\mu_1}{\nu_1}}\phi_{\nu_1}. 
\label{gprimestar}
\end{equation}
($G'^*$ is the ``starred'' function defined \cite{dirac} for $G'$; it 
allows to ``put the star within the bracket'' and to continue with 
the Poisson bracket instead of the Dirac bracket: for any $f$, 
$\{-,\, f \}^{*} = \{-,\, f^{*} \} + pc$) Conditions (\ref{cond2i}) 
and (\ref{cond1i}) are then modified to \begin{equation}
{\partial G'^*\over\partial t} + \{G'^*,H_c\} = pc, \label{cond2j}
\end{equation}
\begin{equation}
\{G'^*,\phi_{\mu_0} \} = B^{\nu_0}_{\mu_0}\phi^1_{\nu_0} + pc, 
\label{cond1j}
\end{equation}
\begin{equation}
\{G'^*,\phi_{\mu_1} \} = pc.
\label{cond3j}
\end{equation}
The Noether transformation $\delta q^{i}$ then becomes \begin{equation} \delta 
q^{i} = {\cal F}\!L^*\{q^{i},\,G'^* \} - v^{\mu_0} ({\cal F}\!L^* 
B^{\nu_0}_{\mu_0}) \gamma^{i}_{\nu_0}. \label{delq2}
\end{equation}

We also learn from this last expression that the conditions for the 
projectability of
the transformation $\delta q^{i}$ are equivalent to the conditions 
for $\delta q^{i}$ to be canonically generated. That is: {\it A 
Noether transformation in phase space is always a canonical 
transformation}. 

Notice that all primary constraints
satisfy (\ref{cond2i}) and (\ref{cond1i}) as does $G$. In the case of 
the primary second class constraints $\phi_{\mu_1}$, $A^{\mu_0}$ and 
$B^{\nu_0}_{\mu_0}$ vanish, and $\delta q^{i}$ in (\ref{delq}) is 
just zero; and so this case is uninteresting. In the case of the 
primary first class constraints $\phi_{\mu_0}$, an interesting case 
is when $\{\phi_{\mu_0}, H_c \} = pc$, that is, when $A^{\mu_0}$ is 
zero. Then $ \epsilon(t)\phi_{\mu_0}$ is a gauge generator for 
arbitrary $ \epsilon$. Other cases
associated with $\phi_{\mu_0}$ must include \cite{cast,kp,gra88,suga} 
a ``chain'' of first class secondary, tertiary, etc., constraints as 
well.

%

\subsection{Geometrical interpretation of the projectability 
conditions}

It is possible to give a geometrical meaning to the projectability 
conditions for
the Noether transformations from tangent space to phase space. 
Observe in (\ref{delq}), or (\ref{delq2}), that it is the presence of 
the matrix
${\bf B} = (B^{\nu_0}_{\mu_0})$ that prevents this transformation 
$\delta q^{i}$ from
being projectable to phase space. Technically, the condition of 
projectability for at least one of the transformations $\delta q^{i}$ 
among the set given by (\ref{arbit}) is that \begin{equation} {\cal 
F}\!L^*(B^{\nu_0}_{\mu_0} \phi^1_{\nu_0})= 0, \ \forall \mu_0, 
\label{projcondit}
\end{equation}
because this makes the $sc$ term in the right side of (\ref{cond1i}), 
or (\ref{cond1j}), a $pc$ term. In such case (\ref{cond2j}), 
(\ref{cond1j}), and (\ref{cond3j}) become \begin{equation}
{\partial G'^*\over\partial t} + \{G'^*,H_c\} = pc, \label{cond2k}
\end{equation}
\begin{equation}
\{G'^*,\phi_{\mu} \} = pc.
\label{cond1k}
\end{equation}
Equations (\ref{cond2k}) and (\ref{cond1k}) are the conditions 
obtained in \cite{b-g-g-p89} to define a projectable Noether 
transformation $\delta q^{i} = {\cal F}\!L^*\{q^{i},\,G'^*\}$. 

There is an elegant way to rephrase the conditions that make the 
Lagrangian Noether transformation projectable to phase space. 
Consider the kernel of the presymplectic form ${\omega}_L$ in tangent 
space. This presymplectic form is defined as the pullback of the 
standard symplectic form in phase space, that is, $$
{\omega}_L := { d}q^i \wedge
{d}\left({\partial L\over\partial\dot q^i}\right). $$
A basis for its kernel is provided \cite{kernel} by the vector fields 
\begin{equation}
{\bf \Gamma}_\mu = \gamma^j_\mu
{\partial\over\partial{\dot q}^j}
\label{nul1}
\end{equation}
and
\begin{equation}
{\bf \Delta}_{\mu_0} = \gamma^j_{\mu_0}
{\partial\over\partial q^j}
+\beta^j_{\mu_0} {\partial\over\partial{\dot q}^j} \ , \label{nul2} 
\end{equation}
with $\beta^j_{\mu_0}$ given by
\begin{equation}
\beta^j_{\mu_0} = K {\partial\phi_{\mu_0}\over\partial p_j} - {\cal 
F}\!L^*\left({\partial\phi^{1}_{\mu_0} \over\partial p_j}\right) \ .
\end{equation}
It is also shown in \cite{kernel}
that, for any function $f(q,p;t)$ on
$T^*Q \times R$, the following property holds: \begin{equation} {\bf 
\Delta}_{\mu_{0}}({\cal F}\!L^*f)
= {\cal F}\!L^*\{f,\phi_{\mu_0}\} \ .
\label{delprop}
\end{equation}

Let us now apply the basis vectors of this kernel to our Noether 
conserved quantity $G_L$. We have (\ref{gammg}),
$$
{\bf\Gamma}_\mu G_L = 0,
$$
and also
\begin{equation}
{\bf \Delta}_{\mu_{0}}G_L =
{\bf \Delta}_{\mu_{0}}({\cal F}\!L^*G)
= {\cal F}\!L^*\{G,\phi_{\mu_0}\} =
{\cal F}\!L^* (B^{\nu_0}_{\mu_0} \phi^1_{\nu_0}), \end{equation} 
where (\ref{delprop}) and (\ref{cond1i}) have been used. Recall that 
the conditions of projectability for $\delta q^{i}$ are equations 
(\ref{projcondit}).
We can write therefore the projectability conditions for $\delta 
q^{i}$ as \begin{equation}
{\bf\Gamma}_\mu G_L = 0, \quad
{\bf \Delta}_{\mu_{0}}G_L = 0.
\label{dq-proj}
\end{equation}

Therefore: {\it The necessary and sufficient condition for a 
Lagrangian Noether conserved function $G_L$ to be associated through 
(\ref{noet}) with a transformation $\delta q^{i}$ that is projectable 
to phase space is that $G_L$ must give zero when acted upon by the 
vector fields in the kernel of the presymplectic form in tangent 
space.}

Notice, as a consequence,
that if all primary constraints are second class (at the primary 
level), then the basis of this kernel is simply ${\bf\Gamma}_\mu$, 
and therefore all Noether transformations may be chosen to be 
projectable.


\subsection{Conditions for the Noether conserved quantity, revisited} 

Equations (\ref{cond1}) and (\ref{cond2}) display the necessary and 
sufficient conditions for a constant of motion $G$ in phase space to 
be associated with a Noether transformation (either projectable or 
not) in tangent space. What is the rationale for these conditions? Of 
course they say that $G$ is indeed a constant of motion. But there 
are constants of motion that only satisfy less restrictive conditions 
that $G$ does, because in (\ref{cond1}) and (\ref{cond2}) use has not 
been made of the full stabilization algorithm --to exhibit all the 
constraints of the theory--, but only the first step. Why is it that, 
for a constant of motion to be Noether, the specific conditions 
(\ref{cond1}) and (\ref{cond2}) must be met? Now we will give an 
independent argument to explain these results.

As we have already seen, the relation
(\ref{noet})
$$
[L]_{i}\delta q^{i} + \frac{d\,G_L}{d\,t} = 0 $$
is telling us that $G_L$ is a constant of motion for the dynamics 
defined by $[L]_{i} = 0$. In (\ref{noet}), $q,\, \dot q, \ddot q$ 
play the role of independent variables. This means that only the 
primary Lagrangian constraints $\chi_\mu=0$
are taken into account, for they are the only constraints 
algebraically included in $[L]_{i} = 0$. The correspondence of these 
constraints with the Hamiltonian constraints has been discussed in 
section 2. The primary Hamiltonian
constraints, $\phi_\mu = 0$, are non-dynamical in the sense that they 
do not appear as a consequence of the equations of motion but only 
define the image in $T^*Q$ of the Legendre map. Recalling section 2 
they classify into first class and second class. The first step of 
the Hamiltonian stabilization algorithm will determine some arbitrary 
functions $\lambda^{\mu_1}$ (the Lagrange multipliers of the primary 
second class constraints) as canonical functions $\lambda_c^{\mu_1}$, 
and some secondary constraints
$\phi^1_{\mu_0} :=\{\phi_{\mu_0},\, H_c \}$. As we have seen in 
(\ref{proj}), the pullback of the secondary constraints gives a 
subset of the primary Lagrangian constraints. The rest of the primary 
Lagrangian constraints, as displayed in (\ref{nonproj}), come from 
comparing the canonical determination $\lambda_c^{\mu_1}$ of the 
functions $\lambda^{\mu_1}$ from their determination $v^{\mu_1}$ in 
tangent space.

It is therefore clear what the status of the Noether canonical 
quantity $G$ must be: It must be a constant of motion for the 
dynamical operator that is obtained after the first step of the 
stabilization algorithm has been performed. We can translate this 
result into
equations. The Hamiltonian dynamics after this first step is given by 
an evolution operator that, using the Dirac bracket, has the form: 
\begin{equation}
{\bf X^1_H} :={\partial \over \partial t} + \{-,\,H_c \}^* + 
\lambda^{\mu_0} \{-,\,\phi_{\mu_0} \} \label{xh1}
\end{equation}
with $\lambda^{\mu_0}$ arbitrary functions. It is irrelevant to write 
a Poisson bracket or a Dirac bracket for the term with 
$\phi_{\mu_0}$. The dynamics given by ${\bf X^1_H}$ is tangent to the 
Hamiltonian primary constraint surface but not necessarily to the 
secondary constraint surface defined by $\phi^1_{\mu_0}= 0$. This is 
parallel to the fact that the Lagrangian dynamics is not necessarily 
tangent to the primary Lagrangian constraint surface, defined by 
$\chi_{\mu}=0$. Since at this stage of the stabilization algorithm we 
have only information on the primary and secondary constraints, under 
the action of the dynamics the constant of motion $G$ must satisfy $$
{\bf X^1_H}(G) =
\frac{\partial G}{\partial t} + \{G,\,H_c \}^* + 
\lambda^{\mu_0}\{G,\,\phi_{\mu_0} \} = sc + pc, $$
but since the functions
$\lambda^{\mu_0}$ are, also at this stage, completely arbitrary, this 
relation splits into
$$
\frac{\partial G}{\partial t} + \{G,\,H_c \}^* = sc + pc, $$
and
$$
\{G,\,\phi_{\mu_0} \} = sc + pc,
$$
which are exactly the conditions (\ref{cond1}) and (\ref{cond2}), 
found in the previous section. We have argued, therefore, that these 
conditions are the correct characterization of the canonical 
constants of motion $G$ associated with Lagrangian Noether 
transformations.

Notice also that in the same way that (\ref{cond1}) and (\ref{cond2}) 
are related to ${\bf X^1_H}$, that is, the dynamics {\it after} the 
first step in the stabilization algorithm, the conditions 
(\ref{cond2k}) and (\ref{cond1k}) that ensure the projectability of a 
Noether transformation are related to the evolution operator ${\bf 
X_H}$ defined in equation (\ref{xh}), that describes the dynamics 
{\it before} the first step in the stabilization algorithm. 

\subsection{The transformation of momenta} 

Up to now, we have been using Hamiltonian techniques to characterize 
the Noether
conserved quantity in phase space, but we have only considered the 
transformation of the
configuration variables, $q$, as in (\ref{delq2}), \begin{equation}
\delta q^{i} = {\cal F}\!L^*\{q^{i},\,G'^* \} - v^{\mu_0} {\cal 
F}\!L^* (B^{\nu_0}_{\mu_0} \{q^{i},\,\phi_{\nu_0}\}). \label{delq3}
\end{equation}
Now we will explore the transformation for the momentum variables. 
Using the tools introduced in section 2, we can compute $\delta{\hat 
p}$, for $\hat p$ defined in ({\ref{hatp}). The result is 
\cite{xgp92} \begin{equation}
\delta{\hat p_{i}} = {\cal F}\!L^*\{p_{i},\,G'^* \} - v^{\mu_0} {\cal 
F}\!L^* (B^{\nu_0}_{\mu_0} \{p_{i},\,\phi_{\nu_0}\}) + [L]_j 
{\partial{\delta q^j}\over\partial{\dot q^{i}}}. \label{delp2} \end{equation}
The last piece vanishes
on shell, that is, when the equations of motion are satisfied, so we 
have an interesting parallelism between (\ref{delq3}) and 
(\ref{delp2}). This new piece $[L]_j {\partial{\delta 
q^j}}/{\partial{\dot q^{i}}}$, will play an important role when we 
consider in section 5 the commutation algebra of transformations in 
configuration space as compared to the Poisson algebra of generators 
in phase space.

Equations (\ref{delq3}) and (\ref{delp2}) suggest an enlargement of 
the formalism: Replace
the functions $v^{\mu_0}(q,\dot q)$ by a set of independent variables 
$\lambda^{\mu_0}$, the Lagrange multipliers, with vanishing Poisson 
bracket with the canonical variables, and define $$
G^c := G'^* - \lambda^{\mu_0} B^{\nu_0}_{\mu_0} \phi_{\nu_0}. $$
Then,
\begin{eqnarray}
\nonumber
\delta q^{i} &=& ({\cal F}\!L^*\{q^{i},\,G^c \})|_{\lambda = v},\\ 
\nonumber
\delta{\hat p_{i}} &=& ({\cal F}\!L^*\{p_{i},\,G^c \})|_{\lambda = v} + 
[L]_j
{\partial{\delta q^j}}/{\partial{\dot q^{i}}}. 
\end{eqnarray}

Since in $\delta{\hat p_{i}}$ the second time derivatives of the 
configuration variables $q^{i}$ are
absorbed within a piece that vanishes on shell, it is natural to 
introduce in this
enlarged space of variables $q$, $p$ and $\lambda$ the Noether 
transformations of the canonical variables as $$
\delta q^{i} =\{q^{i},\,G^c \},\quad
\delta p_{i} =\{p_{i},\,G^c \},
$$
thus putting the transformations of $q$'s and $p$'s on the same 
footing. This is the starting point for the next section.


\section{The enlarged formalism}

It is useful in many respects to reformulate the action principle 
with the canonical Lagrangian,
$$
L_c(q,p,\lambda;\dot q, \dot p, {\dot \lambda}) := p_{i} {\dot q^{i}} 
- H_c(q,p) - \lambda^\mu \phi_\mu(q,p).
$$
The new configuration space for $L_c$ is the old phase space enlarged 
with the Lagrange multipliers $\lambda^\mu$ as new independent 
variables. We use ``enlarged'' instead of ``extended'' to avoid any 
confusion with the ``extended'' Dirac Hamitonian dynamics, where all 
final first class constraints,
primary, secondary, etc., are added to the Hamiltonian with 
independent Lagrange multipliers (in the usual Dirac's theory, which 
is always equivalent to the Lagrangian formulation, only the final 
primary first class constraints are added to the Hamiltonian.) The 
dynamics given by $L_c$ is nothing but the constrained Dirac 
Hamiltonian dynamics for a system with canonical Hamiltonian $H_c$ 
and a number of primary constraints $\phi_\mu$. 

This
formulation has the advantage that all constraints are holonomic, 
that is, defined in configuration space and that the Poisson bracket 
structure defined in the old phase space is still available. Inspired 
by the results of the last section, we will look for Noether 
transformations for $L_c$ that may depend on the Lagrange multipliers 
and their time derivatives to any finite order and that are 
canonically generated for the variables $q^{i}$ and $p_{i}$ 
($\lambda, \dot \lambda, \ddot \lambda, ...$ will have a vanishing 
Poisson bracket with these variables). Let us establish the 
conditions for a function $G^c(q,p,\lambda, \dot \lambda, \ddot 
\lambda,...;t)$ to be a Noether generator, under the definitions
\begin{equation}
\delta_c q ^{i} = \{q^{i}, \, G^c \}, \quad \delta_c p_{i} = \{p_{i}, 
\, G^c \},
\label{delc}
\end{equation}
and with $\delta_c \lambda^\mu$ to be determined below. 

Compute $\delta_c L_c$,
\begin{eqnarray}
\nonumber
\delta_c L_c & = & \delta_c p_i \,{\dot q^i} + {d\over dt}{( p_i 
\,\delta_c q^i)} -
{\dot p_i} \delta_c q^i - \delta_c H_c\\ \nonumber
&-& \lambda^\mu \delta_c \phi_\mu - (\delta_c \lambda^\mu) \phi_\mu \\ 
\nonumber
& = & {d\over dt}{( p_i \,\delta_c q^i)} + \{p_i,\, G^c \} {\dot q^i} 
- \{q^i,\, G^c
\} {\dot p_i}\\ \nonumber 
&-& \{ H_c,\, G^c \}- \lambda^\mu \{ \phi_\mu,\, G^c \} 
- (\delta_c \lambda^\mu) 
\phi_\mu\\ \nonumber
& = & {d\over dt}{( p_i \,\delta_c q^i)} - {\dot q^i} \frac{\partial 
G^c} {\partial q^i} - {\dot p_i} \frac{\partial G^c}{\partial p_i} - 
\{ H_c,\, G^c \}\\ \nonumber
&-& \lambda^\mu \{ \phi_\mu,\, G^c \} - (\delta_c \lambda^\mu) 
\phi_\mu\\ \nonumber
& = & {d\over dt}{( p_i \,\delta_c q^i)} - \frac{d G^c}{d \, t} + 
\frac{\partial G^c}{\partial t}
+ {\dot \lambda^\mu} \frac{\partial G^c}{\partial \lambda^\mu} + 
{\ddot \lambda^\mu} \frac{\partial G^c}{\partial \dot {\lambda^\mu}} 
+ ...\\
\nonumber
&& \, + \, \{ G^c,\, H_c \}
- \lambda^\mu \{ \phi_\mu,\, G^c \} - (\delta_c \lambda^\mu) \phi_\mu 
\\ \nonumber
& = & {d\over dt}{( p_i \,\delta_c q^i - G^c) } + \frac{D G^c}{D t}
+ \{ G^c,\, H_D \}
- (\delta_c \lambda^\mu) \phi_\mu,
\end{eqnarray}
where we have defined the Dirac Hamiltonian \begin{equation}
H_D = H_c + \lambda^\mu \phi_\mu.
\label{hd}
\end{equation}
We have also introduced the notation \cite{htz} 
$$
\frac{D G^c}{D t} :=
\frac{\partial G^c}{\partial t} + {\dot \lambda^\mu} \frac{\partial 
G^c}{\partial \lambda} + {\ddot \lambda^\mu} \frac{\partial 
G^c}{\partial {\dot \lambda}} +... \ ,
$$
and the total time derivative
$$
\frac{d G^c}{d t} :=
\frac{D G^c}{D t} + {\dot q} \frac{\partial G^c}{\partial q} + {\ddot 
q} \frac{\partial G^c}{\partial {\dot q}}. $$
If we require
\begin{equation}
\frac{D G^c}{D t} + \{ G^c,\, H_D \} = pc, \label{enl-cond}
\end{equation}
and if we represent this combination $pc$ of primary constraints as 
$pc = C^\mu \phi_\mu$, then the definition \begin{equation}
\delta_c \lambda^\mu = C^\mu,
\label{dellamb}
\end{equation}
makes
$\delta_c L_c = {d\over dt}{( p \,\delta q - G) }$, that is, a 
Noether transformation for the enlarged formalism. Equation 
(\ref{enl-cond}) is important: it is the equation characterizing the 
generators of Noether transformations in the enlarged formalism. It 
applies both to rigid and gauge Noether symmetries,
and encodes in a compact way the theoretical setting of the results 
given in \cite{htz} (see also \cite{ht}) to find an algorithm to 
produce gauge generators for
theories with only first class constraints. We have arrived at the 
following result: {\it The necessary and sufficient condition for a 
function $G^c(q,p,\lambda, \dot \lambda, \ddot \lambda,...;t)$ to 
generate through (\ref{delc}) a Noether symmetry in the enlarged 
formalism is that $G^c$ must fulfill equation (\ref{enl-cond}).} 

Note that this result has been obtained with no assumptions 
concerning the first and second class structure of the primary 
constraints. Note also that in (\ref{enl-cond}) only the primary 
constraints are relevant.

\subsection{Back to the original Lagrangian} 

We may wonder whether these Noether symmetries for $L_c$ are also 
Noether symmetries for $L$. The answer is yes, and the proof goes as 
follows.

First, we must show how to obtain the original Lagrangian $L$ from 
the canonical Lagrangian $L_c$. Consider the equations of motion 
$[L_c]=0$ for $p$ and $\lambda$,
$$
\dot q^{i} -
{\partial H_c\over\partial p_{i}} + \lambda^\mu {\partial 
\phi_\mu\over\partial p_{i}} = 0,
$$
$$
\phi_\mu = 0.
$$
These equations may be used to obtain $p$ and $\lambda$ in terms of 
the variables $q$ and $\dot q$. We find
$$ p_i = {\hat p_i}(q,\dot q), \quad
\lambda^\mu = v^\mu(q, \dot q),
$$
where the functions $v^\mu$ are those defined in (\ref{id1}) and the 
functions ${\hat p_i}$ are yet to be interpreted. Then the original 
Lagrangian $L$ is retrieved as
$$
L(q,\dot q) = L_c|_{({p_i=\hat p_i}, {\lambda = v})}, $$
and it satisfies that
${\partial L}/{\partial \dot q^i} = \hat p_i$. This is the 
interpretation for the functions ${\hat p_i}$.

What we have just described is the standard method to re-obtain $L$ 
by using the information provided by $H_c$ and the primary 
constraints $\phi_\mu$. This method, when rephrased, as we do here, 
in terms of the canonical Lagrangian,
is just a reduction procedure from $L_c$ to $L$ that can be 
independently justified \cite{ineffective}. We prove in the Appendix 
that it is legitimate to substitute within the Lagrangian ($L_c$ in 
our case) the auxiliary variables, that is, the variables ($p$ and 
$\lambda$ in our case) that can be isolated by using their {\it own} 
equations of motion. 

Now, given a generalized Noether transformation $\delta_c q^{i}$ 
associated, according to (\ref{delc}), with a constant of motion 
$G^c$, we can readily prove that
$\delta q^i := (\delta_c q^i)|_{({p=\hat p}, {\lambda = v})}$ \, 
($\lambda = v$ includes, obviously,
${\dot \lambda = \dot v} =\dot ({\partial v}/{\partial q})+ \ddot 
q({\partial v}/{\partial \dot q})$, and so on for $\ddot \lambda$, 
etc.) defines a Noether transformation for $L$. Indeed we have, by 
definition,
$$
[L_c]_q \delta_c q + [L_c]_p \delta_c p + [L_c]_{\lambda} \delta_c 
{\lambda} + \frac{dG^c}{dt} = 0;
$$
($[L_c]_q$, $[L_c]_p$, and $[L_c]_{\lambda}$ stand for the 
Euler-Lagrange derivatives of $L_{c}$ with respect to the coordinates 
$q^{i}$, $p_{i}$ and $\lambda^{\mu}$, respectively) then, realizing 
that $[L_c]_p|_{({p=\hat p}, {\lambda = v})}$ and 
$[L_c]_{\lambda}|_{({p=\hat p}, {\lambda = v})}$ are identically 
zero, we find
$$
([L_c]_q \delta_c q)|_{({p=\hat p}, {\lambda = v})} + \frac{dG_L}{dt} 
= 0, $$
where $G_L:= G^c|_{({p=\hat p}, {\lambda = v})}$. But, since 
$\lambda$ and $p$ are auxiliary variables for the enlarged formalism, 
we have
$$
[L_c]_q|_{({p=\hat p}, {\lambda = v})} = [L]_q $$
and therefore,
$$
[L]_{i} \delta q^i + \frac{dG_L}{dt} = 0, $$
which proves our assertion.


\subsection{Equivalence between the enlarged and the standard 
formalisms}

First, we will prove that all our results of section 3 can be 
translated to the enlarged formalism. To this end, take $G'^*$ 
satisfying (\ref{cond2j}), (\ref{cond1j}), and (\ref{cond3j}). Then 
the definition
\begin{equation}
G^c := G'^* - \lambda^{\mu_0} B^{\nu_0}_{\mu_0} \phi_{\nu_0} 
\label{gece}
\end{equation}
makes $G^c$ satisfy (\ref{enl-cond}). (The proof is straightforward)

Next we will prove the reverse in the only case for which such a 
proof makes sense, namely, the case of a Noether generator $G^c$ in 
the enlarged formalism that only depends on $\lambda$ but not on its 
time derivatives (in section 3 we considered $\delta q$ and $G_{L}$ 
defined in $TQ\times R$).
Let $G^c(q,p,\lambda;t)$ be such a Noether generator, fulfilling the 
requirements (\ref{enl-cond}). The coefficient of $\dot \lambda$ in 
(\ref{enl-cond}) will therefore satisfy $$
\frac{\partial G^c}{\partial \lambda} = pc, $$
which implies the following form for $G^c$: $$
G^c(q,p,\lambda;t) = G_0(q,p) + G_1^\mu(q,p,\lambda;t) \phi_\mu (q,p).
$$
The Lagrangian conserved quantity $G_L$ is then $$
G_L:= G^c|_{({p=\hat p}, {\lambda = v})} = G_0|_{(p=\hat p)} = {\cal 
F}\!L^* G_0,
$$
and the infinitesimal Noether transformation for $L$ is 
\begin{eqnarray}\nonumber
\delta q^i &:=& (\delta_c q^i)|_{({p=\hat p}, {\lambda = v})} \\ \nonumber
&=& \frac{\partial G_0}{\partial p_{i}}|_{(p=\hat p)} + 
G_1^\mu(q,p,\lambda;t)|_{({p=\hat p}, {\lambda = v})} \gamma_\mu^{i}, 
\end{eqnarray}
which satisfies, according to the results of the previous subsection, 
$[L]_{i} \delta q^i + {d G_L}/{dt} = 0$. 

Since $G_L$ is the pullback of $G_0$, we conclude, owing to the 
results of the previous subsection and those of section 3 (under the 
assumption of the existence of the splitting of the primary 
constraints into first and second class), that $G_0$ satisfies 
(\ref{cond2i}) and (\ref{cond1i}) with some $A^{\nu_0}$ and 
$B^{\nu_0}_{\mu_0}$.
Now, following the same route as we did in section 3.2 from equation 
(\ref{delq}) to (\ref{delq2}), we can define $G'^*_0$, out of $G_0$, 
such that $$
\delta_1 q^i :=
{\cal F}\!L^*\{q^i,\,G'^*_0 \} -
v^{\mu_0} ({\cal F}\!L^* B^{\nu_0}_{\mu_0}) \gamma^{i}_{\nu_0} $$
satisfies $[L]_{i} \delta_1 q^i + {d G_L}/{dt} = 0$. Subtracting this 
equation from the equation satisfied by $\delta q^i$ above gives \begin{equation}
[L]_{i} (\delta q^i - \delta_1 q^i) = 0. \label{del-del}
\end{equation}

The coefficient of $\ddot q^i$ in (\ref{del-del}) tells us that 
$\delta q^i - \delta_1 q^i = s^\mu \gamma_\mu$ for some functions 
$s^\mu$. The rest of the equation
dictates that $s^\mu \chi_\mu = 0$. Thus the difference between 
$\delta q^i$ and $\delta_1 q^i$ is entirely due to the 
indeterminacies already displayed in (\ref{arbit}).

This proves that, except for these irrelevant indeterminacies, that 
are inherent to the formalism, the Noether transformations in the 
tangent bundle
that are obtained through the methods of section 3 are the same than 
those obtained within the
enlarged formalism when we restrict ourselves to dependences on the 
Lagrange multipliers that do not include their time derivatives. 
Moreover, the enlarged
formalism provides us with conditions to be satisfied by a general 
function $G^c(q,p,\lambda, \dot \lambda, \ddot \lambda,...;t)$ in 
order to generate a Noether transformation in the n-th tangent bundle.

\section{The algebra of transformations and generators} 

First consider the simplest case, when the Noether transformations 
are defined in tangent space and are projectable to phase space. This 
means that formulas (\ref{cond2k}) and (\ref{cond1k}) apply in this 
case. Consider two functions
$G_1$ and $G_2$, defined in $T^*Q\times R$, which satisfy, \begin{equation}
{\partial G_r\over\partial t} + \{G_r,H_c\} = pc, \label{cond2kn} \end{equation}
and
\begin{equation}
\{G_r,\phi_{\mu} \} = pc,
\label{cond1kn}
\end{equation}
for $r=1,2$. Then it is straightforward to prove that 
$G_{(2,1)}:=\{G_2,\,G_1\}$ also satisfies the same equations 
(\ref{cond2kn}) and (\ref{cond1kn}). $G_{(2,1)}$ is a Noether 
conserved quantity and a generator of Noether transformations. Does 
$G_{(2,1)}$ generate the commutator transformation $ 
[\delta_1,\,\delta_2]q$? The answer in general is no. Let us 
distinguish between Lagrangian transformations $\delta^L$ and 
Hamiltonian transformations $\delta^H$. If we define $\delta_r^H f:= 
\{f,\,G_r\}$ for any $f \in T^*Q\times R$, then
$$
\delta_r^L q^i = {\cal F}\!L^* (\delta_r^H q^i), $$
and according to (\ref{delp2}),
$$
\delta_r^L{\hat p_{i}} = {\cal F}\!L^* (\delta_r^H p_{i}) + [L]_j 
{\partial{\delta_r^L q^j}\over{\partial{\dot q^i}}}. $$
Now we can compare the Poisson bracket of canonical generators with 
the commutator of Lagrangian transformations. The Jacobi identity 
implies, for the commutator of the Hamiltonian transformations, $$
[\delta_1^H,\,\delta_2^H]q^i = \{q^i,\,\{G_2,\,G_1\}\}. $$
Instead, the commutator of the Lagrangian transformations of the 
configuration variables becomes
\begin{eqnarray}
\nonumber
[\delta_1^L,\,\delta_2^L]q^i = {\cal F}\!L^*(\{q^i,\,\{G_2,\,G_1\}\}) 
\\ 
- [L]_k W_{jl}({{\partial^2 G_1}\over{\partial p_k \partial p_l}}\, 
{{\partial^2 G_2}\over{\partial p_i \partial p_j}}- {{\partial^2 
G_2}\over{\partial p_k \partial p_l}}\, {{\partial^2 
G_1}\over{\partial p_i \partial p_j}}). \label{open} 
\end{eqnarray}

The second term in the right side of equation (\ref{open}) is an 
antisymmetric combination of the
equations of motion. (\ref{open}) displays an open algebra structure 
for the commutator of the
transformations of the configuration variables, that is, an algebra 
that only closes on shell. More specifically, even if we have, at the 
canonical level, a closed Poisson algebra structure, for instance a 
Lie algebra of generators, when we turn to configuration space, the 
commutation
algebra of the transformations will develop a term which is an 
antisymmetric combination of the equations of motion. Unless one of 
the generators involved, $G_1,\ G_2$, is at most linear in the 
momenta, in which case its second partial derivatives in (\ref{open}) 
vanish, these open algebra terms are almost unavoidable. The general 
rule is that {\it a sufficient condition for the commutation 
relations of the projectable Noether transformations in configuration 
space to coincide with the Poisson bracket relations of their 
generators in phase space is that at least all but one of the 
generators are linear in the momenta}. This result was already noted 
in \cite{ba-go-pa-ro89}. An interesting example of a set of 
generators satisfying this sufficient condition is furnished by the 
canonical formulation of general relativity, that is, the ADM 
formalism \cite{adm}.

The gauge algebra plays
a fundamental role concerning the quantization of gauge theories, as 
shown by the BRST methods. In particular, the field-antifield method 
\cite{BV}, that
works in the space of field configurations, is bound to exhibit, in a 
certain number of cases, open algebra structures that originate in 
the type of phenomena here discussed. 

Next, we consider the case of Noether transformations $\delta q$ of 
the type (\ref{delq2}), that is, transformations defined in $TQ\times 
R$ that are not projectable to $T^*Q\times R$. It is convenient to 
work in the framework of the enlarged formalism of section 4, that 
is, we consider
$\delta q^i := (\delta_c q^i)|_{({p=\hat p}, {\lambda = v})}$, with 
$\delta_c q^i = \{q^i, \, G^c \}$, where $G^c$, as defined in 
(\ref{gece}), satisfies (\ref{enl-cond}). If we have two such 
transformations, their corresponding Noether conserved quantities, 
$G^c_1$ and $G^c_2$, will satisfy equation (\ref{enl-cond})
$$
\frac{D \, G^c_r}{D t} + \{ G^c_r,\, H_D \} = C_r^{\mu}\phi_{\mu}, $$
for certain functions $C_r^{\mu}$ ($r=1,2.$). It is easy to check 
that $\{G^c_2,\,G^c_1\}$ does not satisfies (\ref{enl-cond}) in 
general. The reason is that the variation of the Lagrange multipliers 
has not been taken into account. When (\ref{dellamb}) is included, we 
find
$$
G^c_{(2,1)} := \{G^c_2,\,G^c_1\} - (C_1^{\mu_0}B_{2 \mu_0}^{\nu_0} - 
C_2^{\mu_0}B_{1 \mu_0}^{\nu_0}) \phi_{\nu_0}. $$
It is immediate to check that $G^c_{(2,1)}$ satisfies 
(\ref{enl-cond}). Notice that $G^c_{(2,1)}$ contains a new dependence 
in
$\dot {\lambda}^{\nu_0}$ that comes from the quantities 
$C_r^{\mu_0}$. The structure of this
dependence may be displayed by noting that $C_r^{\mu_0}={\dot 
\lambda}^{\nu_0} B_{r\nu_0}^{\mu_0} + ...$, therefore \begin{equation}
G^c_{(2,1)} := \{G^c_2,\,G^c_1\} -
\left({\dot \lambda}^{\nu_0} (B_{1\nu_0}^{\mu_0}B_{2 
\mu_0}^{\sigma_0} - B_{2\nu_0}^{\mu_0}B_{1 \mu_0}^{\sigma_0}) 
+...\right) \phi_{\sigma_0}.
\label{gc21}
\end{equation}

Thus, unless the matrices ${\bf B_1}$ and ${\bf B_2}$ commute, the 
new generator in the enlarged formalism $G^c_{(2,1)}$ will not be 
associated with a transformation of the type (\ref{delq2}) (because 
of this new dependence on ${\dot \lambda}^{\nu_0}$). Now 
$G^c_{(2,1)}$ is a generalized generator of Noether transformations 
defined within the framework of the enlarged formalism of section 4, 
that is, satisfying (\ref{enl-cond}).

Using the two Lagrangian Noether conserved quantities, 
\begin{equation}
G_1^L := {\cal F}\!L^* G^c_1, \quad G_2^L := {\cal F}\!L^* G^c_2, 
\label{gelag}
\end{equation}
we can get the third Lagrangian Noether conserved quantity 
\begin{eqnarray}\nonumber
G^{L}_{(2,1)} &:=& (G^c_{(2,1)}){}_{|_{(p={\hat p},\, \lambda = v)}}\\ 
&=& {\cal F}\!L^*(\{G^c_2,\,G^c_1\}){}_{|_{(\lambda = v)}} \nonumber\\  
&=& {\cal F}\!L^*\{G'^*_2,\,G'^*_1\}\nonumber\\
& -& v^{\nu_0} {\cal 
F}\!L^*(B_{1\nu_0}^{\mu_0}B_{2 \mu_0}^{\sigma_0} - 
B_{2\nu_0}^{\mu_0}B_{1 \mu_0}^{\sigma_0})\chi_{\sigma_0}, 
\label{third}
\end{eqnarray}
but notice that $G^{L}_{(2,1)}$ is not in general projectable to 
phase space because of its dependence on the functions $v^{\nu_0}$ 
(The associated transformation is generated by (\ref{gc21}) and in 
general will exhibit dependences in the accelerations, contained in 
${\dot \lambda}^{\nu_0}$).
Indeed it is projectable if the matrices ${\bf B_1}$ and ${\bf B_2}$ 
commute.


\section{Noether transformations, Rigid and Gauge}

Up to now our discussion has been completely general and it applies 
to both rigid and gauge infinitesimal Noether transformations. Rigid, 
that is, global Noether transformations are physical symmetries of 
the system. Instead, gauge, that is, local Noether transformations 
are unphysical and describe a redundancy of the true degrees of 
freedom of the system \footnote{This sharp statement that 
distinguishes what is physical from what is not, must be refined in 
the context of field theory, because of the possible existence of 
degrees of freedom at the boundaries}. Gauge transformations depend 
on arbitrary functions and are constructed from first class 
constraints. Here we state some considerations and results about both 
types of symmetries. 


\subsection{Noether Rigid transformations} 

The conserved quantity $G$ in $T^*Q \times R$ associated with a rigid 
Noether transformation initially defined in $TQ \times R$ is a 
solution of our general
conditions (\ref{cond1}) and (\ref{cond2}). We know that 
(\ref{cond1}) contains information as to whether the associated 
Noether transformation is projectable to phase space or not. In case 
it is not, in order to get a projectable Noether transformation one 
may try the following. Let us add to $G$ a generator, say $\tilde G$, 
of a gauge transformation where the values of the arbitrary functions 
have been previously fixed as constants. Since the generators of 
gauge transformations are combinations of constraints (see next 
subsection), the value on shell of the conserved
quantity remains unaltered. It is then possible that a convenient 
choice of $\tilde G$ makes the Noether transformation generated by 
$G+\tilde G$ projectable.

We will not discuss the question of the existence of rigid Noether 
symmetries. This problem is closely related with the integrability of 
the dynamical system at hand.
A typical example can be constructed in the case of autonomous 
systems, where the energy is a constant of motion. In field theory, 
the explicit independence of the Lagrangian with respect to the 
space-time variables leads to the existence of a conserved 
energy-momentum tensor. Here also the associated Noether 
transformation can be non-projectable, and the resulting 
energy-momentum tensor may not be gauge invariant. In some cases, 
though, by adding the appropriate gauge generator, it is possible to 
construct a projectable Noether symmetry whose associated 
energy-momentum tensor is gauge invariant \cite{Ja,HenBa}. 

As an example of a rigid transformation that is non-projectable, 
consider the case of an autonomous system with canonical Hamiltonian 
$H_c$. In the place of $G$, $H_c$ itself satisfies (\ref{cond1}) and 
(\ref{cond2}) with $A^{\mu_0} =0$ and
$B^{\mu_0}_{\nu_0} = -\delta^{\mu_0}_{\nu_0}$ (indices $\mu_0$ are 
only available if there are first class constraints at the canonical 
primary level). Hence $H_c$ generates a Noether transformation that 
is not projectable to phase space, an exception being made, of 
course, in the case when all the primary constraints are second class 
among themselves. The fact that the matrix
${\bf B}$ is different from zero prevents the associated $\delta q$ 
from being projectable. Let us be more specific. Considering the 
infinitesimal conserved quantity $ G=\delta t \, H_c$, with $\delta 
t$ an infinitesimal parameter, we have, according to (\ref{delq}), $$
\delta q^j = {\cal F}\!L^*\{q^j,\,\delta t \, H_c \}^* + \delta t \ 
v^{\mu_0} \gamma^{j}_{\mu_0},
$$
where we have used the values for
$A^{\mu_0}$ and $B^{\mu_0}_{\nu_0}$ determined above. Using the 
definition of the Dirac bracket, we can write $\delta q^j$ as \begin{eqnarray}
\delta q^j &=& {\cal F}\!L^*\{q^j, \,\delta t \, H_c \} + \delta t \, 
{\cal F}\!L^* \lambda_c^{\mu_1} \gamma_{\mu_1}^j + \delta t \ 
v^{\mu_0} \gamma_{\mu_0}^j
\nonumber \\
&=& \delta t \left({\cal F}\!L^*\{q^j,\,H_c \} + + v^{\mu_0} 
\gamma_{\mu_0}^j
- M^{{\nu_1}{\mu_1}} \chi_{\nu_1}\gamma_{\mu_1}^j \right) \nonumber \\
&=& \delta t ({\dot q}^j + [L]_i b^{ij}), \end{eqnarray} where we have used the 
identity (\ref{id1}), and where $b^{ij}$ stands for the antisymmetric 
quantity
$b^{ij} := \gamma_{\mu_1}^i M^{{\mu_1}{\nu_1}}\gamma_{\nu_1}^j$. So, 
except for a trivial piece, the transformation $\delta q^j$ is just 
$\delta q^j = \delta t \,{\dot q^j}$, that is, the infinitesimal time 
translation, as it must be.

A lesson may be drawn from this rather simple exploration: {\it The 
existence of a rigid Noether conserved quantity does not guarantee 
that its associated Noether transformation is projectable to the 
canonical formalism}.

In connection with the results of the preceding section, we may also 
note in this example that, since the matrix ${\bf B}$ associated to 
the Hamiltonian is a multiple of the
identity, this matrix will commute with any other matrix ${\bf B_1}$ 
associated with
another non-projectable Noether transformation. That is, the second 
term in the right side of (\ref{gc21}) and (\ref{third}) will vanish. 
In particular, taking $G_1^L$
and $G_2^L$ from (\ref{gelag}), and letting $G_2^L := {\cal F}\!L^* 
H_c$, then it turns out that $G^{L}_{(2,1)}$ in (\ref{third}) is $$
G^{L}_{(2,1)} = {\partial G_1^L \over \partial t}. $$
This
is an expected result for any time independent Lagrangian: The 
explicit time derivative of a conserved Noether quantity is also a 
conserved Noether quantity.


\subsection{Noether Gauge transformations. Existence theorems} 

The generators of Noether gauge transformations are combinations of 
first class constraints (with respect to the final constraint 
surface). They have the general form 
\cite{b-g-g-p89,cast,kp,gra88,suga}
$$
G= \epsilon_{\alpha} G^{\alpha}_{1}+
\dot \epsilon_{\alpha} G^{\alpha}_{2}+\ldots \ , $$
that depends on some combinations $G^{\alpha}_{1},\, G^{\alpha}_{2},\ldots$ of 
first class constraints, and the arbitrary functions $\epsilon_{\alpha}$ 
and their time derivatives. To construct these generators for a given 
theory, it is convenient to solve first the full stabilization 
algorithm, in order to determine the first class constraints needed 
in $G$.

The presence of the
arbitrary functions $\epsilon_{\alpha}$ in the generators of Noether 
gauge transformations makes them very versatile, for we can redefine 
these arbitrary functions with changes of the type
$$
\epsilon_\alpha = f^\beta_\alpha\eta_\beta, $$
with $f^\beta_\alpha(q,p,t)$ a given set of functions, and 
$\eta_\beta$ playing now the role of new arbitrary functions. This 
redefinition of the arbitrary functions amounts to a change of the 
basis of constraints used in the expansion of the generators. The 
usefulness of such changes of basis is twofold: on one side, they 
modify the
algebraic structure of the generators of the gauge group 
\cite{bat-vilk}, and hence one can pass from an ``open'' algebra to a 
``closed'' one, or one can
even end up with an ``abelianized'' algebra. On the other side, they 
may help to make the transformations projectable to phase space. An 
interesting example of this last application is that of the gauge 
group of diffeomorphism-induced transformations in generally 
covariant theories with a metric \cite{pss97}: In order to have these 
transformations projectable to phase space it is compulsory that the 
original arbitrary functions of the spacetime diffeomorphisms include 
some precise dependence on the lapse and shift functions (components 
of the metric in a $3+1$ decomposition).

In general, with regard to gauge transformations, we must distinguish 
the case where all
constraints are first class from the others. In this case, it is 
proven in \cite{ghp} that, under some standard
regularity conditions (constancy of the rank of the Hessian matrix 
and absence of ineffective constraints), Noether canonical gauge 
transformations do exist and with the right number to describe all 
the gauge freedom available to the system, that is, the number of 
final first class primary constraints. As a matter of fact, the proof 
in \cite{ghp} is not completely general but is only valid for cases 
with at most quaternary constraints (three steps of the stabilization 
algorithm being sufficient), but the proof is easily extended to 
cover the general case.

Similar results are obtained within the enlarged formalism of section 
4 for theories with first class constraints and satisfying the same 
regularity conditions. In such case, as proven in \cite{htz}, 
generalized gauge
transformations, depending upon the variables $\lambda, \dot \lambda, 
\ddot \lambda,...$, always exist. The freedom to choose, in this 
case, the basis for the primary constraints, has the price of the 
appearance of dependence on the Lagrange multipliers and their 
derivatives. A careful choice of the basis for the primary 
constraints allows for solutions for the gauge generators independent 
of $\lambda$, in agreement with the results of \cite{ghp}.

In case second class constraints are present, the theorems in 
\cite{ghp} prove that there are still canonical gauge transformations 
that map solutions of the equations of motion into other solutions,
and in the right number, but there is no guarantee that the action is 
conserved up to boundary terms. The difficulty to get a proof for the 
existence of canonical Noether gauge symmetries in this general case 
lies in the structure of the stablilization algorithm, as we now 
discuss.

In \cite{rusos2} the authors claim to have solved in full generality 
the problem of existence of
gauge transformations for theories containing second class 
constraints. In that paper it is taken for granted (formulas (9) and 
(11) of \cite{rusos2}) that a basis for the first class constraints 
exists,
$\Phi_\alpha^{m_\alpha}$; where $\alpha$ numbers the level of the 
stabilization algorithm: primary ($\alpha=1$), secondary 
($\alpha=2$), etc., ${m_\alpha}$ numbers the constraints in the level 
$\alpha$; such that, together with a first class canonical 
Hamiltonian, $H_{FC}$, the following relations hold (the notation is 
the same as in \cite{rusos2,rusos1}):
\begin{eqnarray}
\nonumber
\{\Phi_\alpha^{m_\alpha}, \, H_{FC} \} &=& g_{\alpha \, 
\beta}^{m_\alpha m_\beta} \Phi_\beta^{m_\beta},\\ 
\{\Phi_\alpha^{m_\alpha},\, \Phi_\beta^{m_\beta} \} &=& f_{\alpha \ 
\beta \ \gamma}^{m_\alpha m_\beta m_\gamma} \Phi_\gamma^{m_\gamma}. 
\label{apartheid}
\end{eqnarray}
But this assumption is not proven, neither in \cite{rusos2} nor in a 
preceding paper \cite{rusos1}. In principle one could think that 
(\ref{apartheid}) is a simple consequence of the fact, first proved 
by Dirac \cite{dirac}, that the Poisson bracket between first class 
objects is also first class. This is true, of course, but the 
contents of (\ref{apartheid}) is much more restrictive. Indeed, one 
must take into account that any product of two secondary constraints 
is also first class, and therefore one has, if $\Psi$ stands 
generically for a secondary constraint, \begin{eqnarray}
\nonumber
\{\Phi_\alpha^{m_\alpha}, \, H_{FC} \} &=& g_{\alpha \, 
\beta}^{m_\alpha m_\beta} \Phi_\beta^{m_\beta} + O(\Psi)^2,\\ 
\{\Phi_\alpha^{m_\alpha},\, \Phi_\beta^{m_\beta} \} &=& f_{\alpha \ 
\beta \ \gamma}^{m_\alpha m_\beta m_\gamma} \Phi_\gamma^{m_\gamma} + 
O(\Psi)^2,
\end{eqnarray}
where $O(\Psi)^2$ stands for any piece quadratic in the secondary 
constraints. It is not difficult to get rid of these quadratic pieces 
for the Poisson bracket of the $\alpha$-level first class constraints 
with the first class Hamiltonian, simply by defining the $(\alpha + 
1)$-level first class constraints as the results of these Poisson 
bracket and disregarding the redundant constraints that may result. 
But then we cannot prevent the Poisson bracket of first class 
constraints from developing quadratic
pieces in the secondary constraints. Or vice-versa, we can write the 
constraints, if they are all effective, in a new basis such that all 
are canonical variables, a ``Darboux'' basis. In such case, the 
Poisson bracket between first class constraints has no quadratic 
pieces -it just vanishes- but nothing prevents these quadratic pieces 
from being present in the Poisson bracket of the first class 
constraints with the first class Hamiltonian. 

Since the assumption (\ref{apartheid}) seems instrumental in 
obtaining the existence theorems for Noether gauge transformations, 
we must therefore assert that there has not yet been produced a 
general proof of the existence of canonical Noether gauge symmetries 
for systems with first as well as second class constraints. 
Nevertheless, we are going to prove in the next subsection a theorem 
of existence for canonical Noether gauge transformations for general 
theories having only primary and secondary constraints, that is, 
theories whose stabilization algorithm has only one step. Since most 
of the physical cases, like general
relativity or Yang-Mills theories, fall into this case, we can say 
that our results, though incomplete, do have some interest.


\subsection{Existence of canonical Noether gauge transformations for 
theories with one-step stabilization algorithm} 

Consider a theory satisfying our standard regularity conditions, with 
canonical Hamiltonian $H_c$, and with a set of primary constraints. 
We classify them into first and second class. Next we look for the 
secondary constraints, which we will suppose to be effective. They 
introduce new restrictions to the primary
constraint surface, so that they are defined up to the addition of 
primary constraints. And suppose, also, that there are no more levels 
(tertiary,...) of constraints. Some of the secondary constraints make 
second class some of the former first class primary constraints. The 
rest of the secondary constraints, chosen in a convenient basis, will 
be first class. By changing the basis for the primary and secondary 
constraints and performing a subsequent canonical transformation, we 
can express all the constraints in a ``Darboux'' basis, 
\begin{eqnarray}\nonumber
Primary &:& \, P_1,...,P_m,\, P_{m+1},...,P_n,\, 
P_{n+1},...,P_{n+r},\\ \nonumber
&&Q_{n+1},...,Q_{n+r}.\\ 
\nonumber
Secondary &:& \, 
P'_1,...,P'_l,\, Q_{m+1},...,Q_n. 
\end{eqnarray}
$P_1,...,P_m$ are the final first class primary constraints, 
$P_{m+1},...,P_n$ are the former first class primary constraints that 
become second class when the secondary constraints are introduced. 
$P_{n+1},...,P_{n+r},\,Q_{n+1},...,Q_{n+r}$ are couples of canonical 
variables corresponding to the primary second class constraints. 
$P'_1,...,P'_l$, with $l \leq m$, are the secondary first class 
constraints, and $Q_{m+1},...,Q_n$ are the secondary second class 
constraints that make the primary subset $P_{m+1},...,P_n$ second 
class.

Since the canonical Hamiltonian is only determined up to primary 
constraints, we can, without any lost of generality, set to zero in 
$H_c$ the variables corresponding to the primary constraints. Notice 
that, since the Poisson bracket between first class objects is also 
first class, we will have, for this new $H_c$, $$
\{P_i, \, H_c \} = A^a_i P'_a + D_i^{st}Q_sQ_t, $$
for some matrices (of functions) $A^a_i$ and $D_i^{st}$; $\ i= 
1,...,m$, $\ a=1,...,l$,
$\ s,t = m+1,...,n$.
$A^a_i$ is a maximum rank matrix. If $l < m$, $A^a_i$ has $m-l$ null 
vectors $C^i_\sigma$, $\ C^i_\sigma A^a_i = 0$, $\ \sigma = 
l+1,...,m$.

Having done all these preliminaries,
finding $m$ independent canonical Noether gauge quantities enclosed 
in $G$, $G= \epsilon^i(t) G_i, \, i=1,...,m.$, with $\epsilon^i$ 
arbitrary functions, becomes trivial. They are, $$
G_a = P'_a, \, a=1,...l; \qquad G_\sigma = C^i_\sigma P_i, \, \sigma 
= l+1,...,m.
$$

It is trivial to check (\ref{cond2i}) and (\ref{cond1i}) for all 
these quantities. The coefficients $A^{\mu_0}$ in (\ref{cond2i}) 
vanish for $G_\sigma$ but not
necessarily for $G_a$. Since $B^{\nu_0}_{\mu_0}=0$ for all $G_a$, 
$G_\sigma$, we
know by (\ref{dq-proj}) that their associated Noether transformations 
are canonical.

Thus we have arrived at the following result: {\it Any theory with 
only primary and secondary constraints (all effective) exhibits a 
basis of independent canonical Noether gauge generators in a number 
that equals the number of final primary first class constraints} 

This proof of existence of Noether gauge transformations cannot be 
easily generalized to theories whose
stabilization algorithm has more than one step. Of course one can 
prepare the constraints in a ``Darboux'' basis, where conditions 
(\ref{cond1i}) can be readily met. But the unsolved problem is to 
find the right number of objects, combinations of first class 
constraints, satisfying (\ref{cond2i}). This is still an open problem.


\section{Examples}

\subsection{Example 1: Dirac Hamiltonian} 

For any time independent first order Lagrangian, the Dirac 
Hamiltonian $H_D$ (\ref{hd}), which is also time independent, 
satisfies the condition (\ref{enl-cond}) to be a generator of Noether 
transformations. These transformations, $\delta_c q = \{ q,\, \delta 
t H_D \}$ become, in tangent space,
\begin{eqnarray}
\nonumber
\delta q^i &:=& (\delta_c q^i)|_{({p=\hat p}, {\lambda = v})}\\
\nonumber
& =& \delta 
t ({\cal F}\!L^* \{ q^i,\,H_c \} + v^\mu {\cal F}\!L^*\{ q^i,\, 
\phi_\mu \})\\ \nonumber
& =& \delta t (K q^i) = \dot q^i \delta t. 
\end{eqnarray} 
This is an 
expected result that was already discussed in the context of 
non-projectable rigid Noether symmetries in subsection 6.1 

\subsection{Example 2: Presence of terms quadratic in the 
constraints}

Having theorems that guarantee the existence of gauge Noether 
transformations is not the same as finding them in practice. Writing 
the constraints in a ``Darboux'' basis may prove cumbersome, and in 
many cases it is advantageous to circumvent these procedures and to 
obtain the Noether conserved quantities from simpler considerations. 
Here we consider an example exhibiting first and second class 
constraints. It illustrates some special features that are absent in 
the case with only first class constraints. Our Hamiltonian is
$$
H_c = \frac{1}{2} (p_1^2 + p_2^2),
$$
where $p_1$ and $p_2$ are vectors in Minkowski space. The primary 
constraints are the following scalar products $$
\phi_1 = (p_1,x_2) = 0,\qquad \phi_2 = (p_2,x_2) = 0, $$ where $x_2$ 
is the vector whose components are the canonical coordinates 
conjugate to those of $p_2$. The corresponding Lagrangian is
$$
L = \frac{1}{2}({\dot x}_1^2 - \frac{({\dot x_1}x_2)^2}{x_2^2}) + 
\frac{1}{2}({\dot x}_2^2 - \frac{({\dot x_2}x_2)^2}{x_2^2}). $$ 

Both constraints are first class on the primary surface, and 
$\{\phi_1,\, \phi_2\} = \phi_1$. Their stabilization gives the 
secondary constraints:
$$\xi_1 = \{\phi_1,\, H_c \}= (p_1,p_2), \qquad \xi_2 = \{\phi_2,\, 
H_c \}= p_2^2.
$$
No more constraints appear. Notice that $\{\phi_1,\, \xi_1 \} = 
p_1^2$. Therefore, if
we restrict ourselves to a region in phase space with $p_1^2 \neq 0$, 
the constraints $\phi_1$ and $\xi_1$ become second class. For a 
correct definition of $L$, we will also assume that our region 
satisfies $x_2^2 \neq 0$.

The obvious candidate for a conserved quantity $G$ associated with a 
Noether gauge transformation is the secondary constraint $\xi_2$, 
which is a final first class constraint. In fact, $G = \epsilon(t) 
\xi_2$, with $\epsilon(t)$ an arbitrary function, satisfies 
(\ref{cond2i}) and (\ref{cond1i}) with $A^1 = A^2 = 0$ and $B_1^1 = 
B_2^2 = -2$. Thus, we have at hand the identification of the Noether 
transformation $\delta q$ associated with $G$, just by using 
(\ref{delq}). Since ${\cal F}\!L^* B^{\nu_0}_{\mu_0} \neq 0$, we know 
by (\ref{dq-proj}) that $\delta q$ is not projectable. Our experience 
with theories with only first class constraints tells us that to get 
a projectable $\delta q$ we may try to replace $\epsilon(t)$ by a 
function $\epsilon = \eta(t) f(q,p)$ with $\eta$ arbitrary and $f$ to 
be determined so as to render $B^{\nu_0}_{\mu_0} = 0$. This 
substitution amounts to a change of basis for the secondary first 
class constraint. But this substitution does not works in our case. 
The subtle point is that the change of basis $\xi_2 \rightarrow 
f\xi_2$ is not general enough, that is, we have not fully used the 
freedom to modify $G$. In fact, the square of the secondary second 
class
constraint $\xi_1$ is first class( because any ineffective constraint 
is). Therefore, we may add to $G$ a term linear in $(\xi_1)^2$,that 
is, we can take $G = \eta(t) (f \xi_2 + g (\xi_1)^2)$. Now we can 
find simple solutions for $f$ and $g$ that make $B^{\nu_0}_{\mu_0} = 
0$, for instance $$
G = \eta(t) (x_2^2 p_1^2 \xi_2 - x_2^2(\xi_1)^2). $$

The values of $A^1$ and $A^2$, according to (\ref{cond2i}), for this 
$G$, may be used to obtain, by means of (\ref{delq2}), the canonical 
generator $G'^*$ of a projectable $\delta q$. We get $$
G'^* = \eta(t) (x_2^2 p_1^2 \xi_2 - x_2^2(\xi_1)^2) + {\dot \eta}(t) 
(x_2^2 p_1^2 \phi_2 - x_2^2 \xi_1 \phi_1). $$
Notice that all terms use constraints that are first class, though 
some are the product
of final second class constraints. Had we not considered these 
quadratic pieces, we would have not a canonical generator of a 
projectable Noether gauge transformation. Obviously these special 
features disappear if we use a ``Darboux'' basis for the constraints, 
but it is harder to get this basis than to proceed along our lines. 


\subsection{Example 3: A non-projectable Noether transformation} 

As an example which exhibits a non-projectable Noether 
transformation, we take the Lagrangian
\begin{equation}
L=\frac12 {\dot q}_1^2+\frac{1}{2q_4}{\dot q}_2^2+\frac12 {\dot 
q}^2_3 +\frac12 q^2_2 + q_3q_5,
\end{equation}
which differs from the example presented in \cite{rusos2} by a simple 
canonical transformation.
The primary constraints are $p_4=0$ and
$p_5=0$.
The canonical Hamiltonian is
\begin{equation}
H_c=\frac12 p^2_1+\frac12 q_4p_2^2+\frac12 p_3^2 -\frac12 q^2_2 
-q_3q_5
\end{equation}
The stabilization of $p_4=0$ give rise to the chain of first class 
constraints
$$
\phi_1 =p_4,\, \phi_2 =-\frac12 p^2_2,\, \phi_3 =-q_2p_2, \, 
\phi_4=-q_2^2-q_4p_2^2,
$$
while the stabilization of $p_5=0$ only gives rise to second class 
constraints:
$$
\psi_1=p_5, \psi_2=q_3,
\psi_3=p_3, \psi_4= q_5.
$$
Using the Dirac bracket we can eliminate the canonical pairs 
$(q_3,p_3)$ and $(q_5,p_5)$ though it is not necessary. A solution to 
our
fundamental conditions (\ref{cond2i}) and (\ref{cond1i}) is 
\begin{equation}
G=\phi_2 (\ddot\epsilon -4 q_4\epsilon) -\phi_3\dot \epsilon 
+\phi_4\epsilon,
\end{equation}
with $A={\buildrel \ldots\over\epsilon} - 4q_4\dot \epsilon$ 
and $B=-4\epsilon$.  Since
$B\neq 0$ the generator has an associated non-projectable Noether 
symmetry. It is easy to show that in this case it is not
possible to use a redefinition of the constraint surface to render 
the corresponding transformation projectable. Notice that, in 
particular, the
transformation $\delta q_4=- A+4{\dot q}_4 \epsilon$ is not 
canonically generated. Nevertheless, we can use the enlarged 
formalism of section 4 to define a canonical generator $G_c$ of this 
non-projectable Noether symmetry. To accomplish this we promote the 
Lagrange multiplier $\lambda$ associated to the primary first class 
constraint $p_4=0$ to a dynamical variable --- that is, we work in 
the enlarged formalism --- and use the relation (\ref{gece}) to 
define a canonical generator \begin{equation}
G^c=G+(4\lambda\epsilon-A)p_4.
\end{equation}
This canonical generator satisfies equation (\ref{enl-cond}) ( the 
Dirac Hamiltonian is $H_D=H_c+\lambda p_4$. The canonical 
transformation $\delta_c q_4$ generated by $G^c$, $\delta_c 
q_4=4\epsilon\lambda
-A$, coincides with the previous non-canonical symmetry upon the 
substitution $\lambda={\dot q}_4, p_4=0$ that comes from the 
equations of motion for $p_4$ and $\lambda$, respectively. This is an 
expected result according to the general arguments introduced in 
section 4.

\subsection{Example 4: Abelian Chern-Simons field theory in $2n+1$ 
dimensions}

Here we present, as a field theory example, the generic Abelian 
Chern-Simons theory in $2n+1$ dimensions \cite{FP,bana96}. We use 
some results of \cite{bana96} and, mostly, their notation. Our 
treatment differs from \cite{bana96} in that we retain all the 
variables, including $A_{0}$.

Abelian Chern-Simons theory has only
primary and secondary constraints; therefore, in principle, a Noether 
gauge generator can be constructed. One must be aware that in some 
field theories the basis of constraints to achieve this goal can be 
involved. Indeed, there are
cases where the Noether generator is a non-local function, as in the 
canonical formulation of the electromagnetic duality transformation 
(see for example \cite{Be}). In our case, however, our procedures 
will not meet these difficulties.

The Lagrangian density is
\begin{equation}
{\cal L}_{CS}^{2n+1}= \epsilon^{\mu_{0}
\mu_1...\mu_{2n}}A_{\mu_{0}}F_{\mu_{1}\mu_{2}} \ldots 
F_{\mu_{2n-1}\mu_{2n}}, \label{cslagr}
\end{equation}
where the greek indices $\mu$ run from $0$ to $2n$, $F_{\mu\nu} 
:=\partial_{\mu} A_{\nu}- \partial_{\nu} A_{\mu}$ and 
$\epsilon^{\mu_{0} \mu_1...\mu_{2n}}$ is totally antisymmetric with 
$\epsilon^{012\ldots,2n}=1$.
The Noether gauge invariance for ${\cal L}_{CS}^{2n+1}$ is given by 
\begin{equation}
\delta A_{\mu} = - \partial_{\mu}\epsilon + F_{\mu \nu} \eta^{\nu}, 
\label{csgauge}
\end{equation}
where $\epsilon$ and $\eta^{\mu}$ are infinitesimal arbitrary 
functions (the usual diffeomorphisms are recovered by taking 
$\epsilon = A_{\nu} \eta^{\nu}$).

Eventhough there is no
metric, we take, as it is customary, the $x^0$ coordinate as the time 
evolution parameter. Since ${\cal L}_{CS}^{2n+1}$ does not
depend on $\dot A_{0}$, it is convenient to identify in 
(\ref{cslagr}) the terms containing $A_{0}$. The Lagrangian is 
equivalently written as (the latin indices $i$ run from $1$ to $2n$) 
\begin{equation}
{\cal L}_{CS}^{2n+1}= A_{0} K + (\partial_{0}A_{i}- 
\partial_{i}A_{0}) l^{i},
\end{equation}
with
$$
K := \epsilon^{0i_1...i_{2n}}F_{i_{1}i_{2}} \ldots 
F_{i_{2n-1}i_{2n}}, $$
and
$$
l^{i} := 2n \epsilon^{0iji_{3}...i_{2n}}A_{j}F_{i_{3}i_{4}} \ldots 
F_{i_{2n-1}i_{2n}}.
$$
(Notice that $\partial_{i}l^{i} = n K$) We can go to the Hamiltonian 
formulation. The Lagrangian momenta are
\begin{eqnarray}
\hat p^{0} &:=& {\partial {\cal L}_{CS}^{2n+1} \over \partial {\dot 
A_{0}}} = 0, \nonumber \\
\hat p^{i} &:=& {\partial {\cal L}_{CS}^{2n+1} \over \partial {\dot 
A_{i}}} = l^{i}, \nonumber
\end{eqnarray}
whereby we can read off the primary Hamiltonian constraints \begin{equation}
\phi^{0} := p^{0} = 0, \qquad
\phi^{i} := p^{i} - l^{i} = 0.
\end{equation}
(These equalities to zero are weak equalities in the Dirac sense: 
they are part of the equations of motion)
The canonical Hamiltonian is determined only up to terms linear in 
the primary constraints. A simple choice is
\begin{equation}
H_{c} = -A_{0} (K + \partial_{i}p^{i}).
\end{equation}
(This Hamiltonian differs from the one taken in \cite{bana96} by 
terms linear in the primary constraints). At this point, the dynamics 
is determined by the vector field in phase space \begin{equation}
{\bf X_{H}} = {\partial \over \partial t} + \{ -, \, H_{c} \} + 
\lambda_{\mu}\{ -, \, \phi^{\mu}\}.
\end{equation}

Let us obtain, for future use, the Lagrangian determination $v_{\mu}$ 
of the Lagrange multipliers $\lambda_{\mu}$, as discussed in section 
2. Since the time derivatives of the configuration variables 
$A_{\mu}$ are $$
{\dot A_{0}} = {\bf X_{H}}(A_{0}) = \lambda_{0}, \qquad
{\dot A_{i}} = {\bf X_{H}}(A_{i}) = \partial_{i}A_{0} + \lambda_{i}, 
$$
we find
\begin{equation}
v_{0}= {\dot A_{0}}, \qquad v_{i}= F_{0i}. \label{visf}
\end{equation}
These functions $v_{0},\, v_{i}$ are clearly non-projectable to phase 
space because the Hessian matrix for ${\cal L}_{CS}^{2n+1}$ vanishes 
identically (the Lagrangian is linear in the velocities) and 
therefore the vectors $\Gamma_{\mu}$ in (\ref{GAMMA}) are now 
$\Gamma_{\mu} = {\partial \over \partial{\dot A_{\mu}}}$.

We have the following Poisson brackets:
\begin{eqnarray}
\{\phi^{0}, \, \phi^{i} \} &=& 0, \nonumber \\ \{\phi^{i}, \, 
\phi^{j} \} &=& - 2n(n+1) \epsilon^{0iji_{3}...i_{2n}}F_{i_{3}i_{4}} 
\ldots F_{i_{2n-1}i_{2n}} =: \Omega^{ij} \nonumber \\ \{\phi^{0}, \, 
H_{c} \} &=& K + \partial_{i}p^{i} =: \psi^{0} \nonumber \\
\{\phi^{i}, \, H_{c} \} &=& 0.
\label{thepb}
\end{eqnarray}

The stabilization algorithm for the primary constraint $\phi^{0}= 0$ 
gives the secondary constraint $\psi^{0} =0$. The other primary 
constraints give
\begin{equation}
0 ={\dot \phi^{i}} ={\bf X_{H}}\phi^{i} = \Omega^{ij} \lambda_{j}. 
\label{stabphi}
\end{equation}
To continue, it is necessary to know the rank of the matrix 
$\Omega^{ij}$. The identity
\begin{equation}
\Omega^{ij} F_{jk} = (n+1) \delta^{i}_{k} K, \label{omegaefa}
\end{equation}
is telling us, taking into account that
$$
(n+1) K = \psi^{0} - \partial_{i}\phi^{i}, $$
that neither $\Omega^{ij}$ nor $F_{jk}$ can have rank $2n$, on shell.
From now on we will work in the region of configuration space where 
the theory is generic \cite{bana96}, that is, $F_{ik}$ has the 
maximum rank compatible with the equations of motion. This maximum 
rank is $2n-2$. We take \cite{bana96} coordinates $x^{i} = 
x^{\alpha},\, x^{p}; \, \alpha = 1, 2; \, p = 3,\ldots,2n.$, such 
that the maximum rank condition is attained by $F_{pq}$, that is, 
$\det(F_{pq}) \neq 0$. 

Define $F^{pq}$ as the matrix inverse to $F_{pq}$. Equation 
(\ref{omegaefa}) implies the on shell (that is, when $K=0$) 
relationship $$
\Omega^{i p}= -\Omega^{i \beta}F_{\beta q}F^{qp}, $$
that shows that the maximum rank for $\Omega^{ij}$ is $2$, on shell. 
Indeed this rank is attained by $\Omega^{\alpha\beta}$. Defining its 
inverse matrix as $\Omega_{\alpha\beta}$, we can write, using (\ref 
{omegaefa}), the identities
\begin{eqnarray}\nonumber
F_{\alpha q}F^{qp} + \Omega_{\alpha\beta}\Omega^{\beta p} &=& 0, \\ 
\Omega^{pq} - \Omega^{p\alpha}\Omega_{\alpha\beta}\Omega^{\beta q} &=& 
(n+1) F^{pq}K
\label{nice}
\end{eqnarray}
that will be used below.

The preceding analysis shows that, being the theory generic, the only 
primary second class constraints are $\phi^{1}$ and $\phi^{2}$. The 
Dirac bracket is then defined as
$$
\{-,\, -\}^{*} = \{-,\, -\} -
\{-,\, \phi^{\alpha}\}\Omega_{{\alpha \beta}}\{\phi^{\beta},\, -\}. $$

The Lagrangian multipliers $\lambda_{1}$ and $\lambda_{2}$ become 
determined by (\ref{stabphi}). Recalling (\ref{xh1}), the new 
evolution operator is
$$
{\bf X_{H}^{1}} = {\partial \over \partial t} + \{ -, \, H_{c} \}^* + 
\lambda_{p}\{ -, \, \phi^{p}\}^{*} + \lambda_{0}\{ -, \, 
\phi^{0}\}^{*}. $$

Our next task is to find the Noether gauge generators. Our 
Chern-Simons theory exhibits a special circumstance which is worth 
mentioning: In contrast
to the standard case developed in section 3, our primary first class 
constraints only close (to exhibit their first class condition) when 
the secondary constraint is also taken into account. This is due to 
the fact that in the determination of the rank of $\Omega^{ij}$ (in 
the generic case) use has been made of the secondary constraint. 
Nevertheless, since the closure of the primary first class 
constraints only involves the secondary constraint our theory still 
applies, as we will see.
Using the second identity in (\ref{nice}) we compute this closure 
property. It reads:
\begin{equation}
\{\phi^{p}, \, \phi^{q}\}^{*} = (n+1) F^{pq}K = F^{pq}\psi^{0} + pc. 
\label{closure}
\end{equation}

Now we can look for the gauge generators. It is clear that 
$\epsilon_{0}\psi^{0}$ and
$\epsilon_{p}\phi^{p}$, with $\epsilon_{0},\, \epsilon_{p}$ arbitrary 
functions, satisfy the requirements for the function $G$ in 
$(\ref{cond2i})$ and $(\ref{cond1i})$. For $\epsilon_{0}\psi^{0}$ the 
coefficient $A$ in $(\ref{cond2i})$ is $\dot \epsilon_{0}$ and the 
coefficient $B$ in $(\ref{cond1i})$ vanishes. For 
$\epsilon_{p}\phi^{p}$ the coefficient $A$ vanishes and $B^{q}_{0} = 
\epsilon_{p}F^{pq}$. Thus the transformations generated by 
$\epsilon_{p}\phi^{p}$ are not projectable. This non-projectability 
suggests us to work in the enlarged formalism. Let $$
G := \epsilon_{0}\psi^{0} + \epsilon_{p}\phi^{p}, $$
then, the quantity $G'^{*}$ in (\ref{gprimestar}) becomes 
($\phi^{0*}=\phi^{0}, \, \psi^{0*}=\psi^{0} $): \begin{eqnarray}
G'^{*} &=& \epsilon_{0}\psi^{0} -{\dot \epsilon_{0}}\phi^{0}+ 
\epsilon_{p}\phi^{p*} \nonumber \\
&=& \epsilon_{0}\psi^{0} -{\dot \epsilon_{0}}\phi^{0} + 
\epsilon_{p}(\phi^{p} - \Omega^{p\alpha} \Omega_{\alpha 
\beta}\phi^{\beta}) \nonumber \\ &=& \epsilon_{0}\psi^{0} -{\dot 
\epsilon_{0}}\phi^{0} + \epsilon_{p}(\phi^{p} + F_{\beta 
q}F^{qp}\phi^{\beta}),
\end{eqnarray}
where we have used the first identity in (\ref{nice}). Finally, the 
generator $G^{c}$ in (\ref{gece}) is
\begin{equation}
G^{c} = \epsilon_{0}\psi^{0}-{\dot \epsilon_{0}}\phi^{0}+ 
\epsilon_{p}(\phi^{p} +
F_{\beta q}F^{qp}\phi^{\beta} + \lambda_{q}F^{qp}\phi^{0}), \end{equation}
and it is straightforward to prove that $G^{c}$ satisfies equation 
(\ref{enl-cond}). It is convenient to redefine the arbitrary 
functions $\epsilon_{p}$ as $\epsilon_{p}= F_{pq}\eta^{q}$. Then \begin{equation}
G^{c} = \epsilon_{0}\psi^{0} -{\dot \epsilon_{0}}\phi^{0} + 
\eta^{q}(F_{pq}\phi^{p} + F_{\beta q}\phi^{\beta} + 
\lambda_{q}\phi^{0}), \end{equation}
Recalling the Lagrangian determination (\ref{visf}) for 
$\lambda_{q}$, we can write the gauge transformations generated by 
$G^{c}$, (\ref{delc}),
\begin{eqnarray}
\delta A_{0} &=& \{A_{0},\, G^{c}\}{}_{|_{(\lambda=v)}} = - {\dot 
\epsilon_{0}} + F_{0q}\eta^{q} \nonumber \\ \delta A_{\beta} &=& 
\{A_{\beta},\, G^{c}\}{}_{|_{(\lambda=v)}} = - {\partial_{\beta} 
\epsilon_{0}} + F_{\beta q}\eta^{q} \nonumber \\ \delta A_{p} &=& 
\{A_{p},\, G^{c}\}{}_{|_{(\lambda=v)}} = - {\partial_{p} 
\epsilon_{0}} + F_{pq}\eta^{q}, \end{eqnarray}
or, in a more compact way,
\begin{equation}
\delta A_{\mu} = - {\partial_{\mu} \epsilon_{0}} + F_{\mu q}\eta^{q}. 
\label{thegaugetr}
\end{equation}

The time diffeomorphism and the diffeomorphims in the $x^{1}$ and 
$x^2$ directions are hidden in (\ref{thegaugetr}). To find them, one 
must consider the Lagrangian
equations of motion,
$$
\Omega^{ij}F_{0j} = 0.
$$
(They can be read off from (\ref{stabphi}) after using the lagrangian 
determination (\ref{visf}) for $\lambda^{j}$). Since a basis for the 
null vectors for $\Omega^{ij}$ is already spanned by $F_{jp},\, p = 
3,4,\ldots,2n.$, we conclude that there exist matrices 
$N^{q}_{\alpha}, \,N^{q}_{0}$ such that, on shell, \begin{equation}
F_{\alpha\mu} = N^{q}_{\alpha}F_{q\mu }, \qquad F_{ 0\mu} = 
N^{q}_{0}F_{ q\mu}.
\label{reducible}
\end{equation}
Then, since $\eta^{q}$
are arbitrary functions, they can be expressed as $$\eta^{q} = 
\rho^{q} + N^{q}_{\alpha}\rho^{\alpha} + N^{q}_{0}\rho^{0},
$$
with $\rho^{q}, \,\rho^{\alpha}, \,\rho^{0}$ arbitrary functions. 
Therefore, on shell
\begin{equation}
\delta A_{\mu} = - {\partial_{\mu} \epsilon_{0}} + F_{\mu 
\nu}\rho^{\nu}. \label{thegaugetr2}
\end{equation}
We have recovered the complete set of gauge transformations 
(\ref{csgauge}) that in fact are true Noether transformations, on and 
off shell. Obviously they are not all independent: the time and the 
$x^{1}, \, x^{2}$ diffeomorphisms can be obtained either 
by using the arbitrary functions $\rho^{\alpha}, \, \rho^{0}$ or as a 
byproduct of the $x^{p}$ diffemorphisms, as we have just seen. The 
reason for this is the fact that the dynamics introduces relations 
between the components of the curvature tensor $F_{\mu\nu}$ that 
produce a redundancy in the action on shell of the algebra of 
diffeomorphisms. Thus, the gauge algebra is reducible on shell. This 
is a well known fact that, in the language of BRST methods 
\cite{quimparis}, implies the introduction of ``ghosts for ghosts'' 
in the formalism. Indeed, the transformations (\ref{thegaugetr2}) are 
reducible on shell because, if we take \begin{equation}
\delta A_{\mu} = R_{\mu \nu}\rho^{\nu},
\label{erragran}
\end{equation}
with $R_{\mu \nu} = F_{\mu \nu}$, then there exist matrices 
$Z^{\nu}_{\alpha},\, Z^{\nu}_{0}$, with
$$
Z^{\nu}_{\alpha} = (0,\,\delta _{\alpha}^{\beta}, \, - 
N_{\alpha}^{p}), \qquad
Z^{\nu}_{0} = (1,\,0, \, - N_{0}^{p})
$$
such that, according to (\ref{reducible}), the following relations 
hold on shell,
$$
Z^{\nu}_{\alpha}R_{\mu \nu}=0, \qquad Z^{\nu}_{0}R_{\mu \nu}=0. $$

This redundancy has been expressed in a non covariant form. A full 
covariant treatment can be given by noticing that the covariant form 
of the equations of motion (including the constraint $K=0$) is $$ 
Q^{\mu} := \epsilon^{\mu
\mu_1...\mu_{2n}} F_{\mu_{1}\mu_{2}} \ldots F_{\mu_{2n-1}\mu_{2n}} 
=0. $$
( $Q^{0} =0$ is $K=0$, $Q^{i}=0$ is
$\Omega^{ij}F_{0j} = 0$) If we define
$$
\Omega^{\mu\nu\rho}:= \epsilon^{\mu \nu\rho \mu_3\mu_4...\mu_{2n}} 
F_{\mu_{3}\mu_{4}} \ldots F_{\mu_{2n-1}\mu_{2n}}, $$
the following identity holds:
$$
\Omega^{\mu\nu\rho}F_{\rho\sigma}= {1\over 
2n}(\delta^{\mu}_{\sigma}Q^{\nu}- \delta^{\nu}_{\sigma}Q^{\mu}). $$
Then, the covariant expression for the reducibility of the gauge 
transformations (\ref{erragran}) is expressed, on shell, as $$
\Omega^{\mu\nu\rho}R_{\sigma\rho} =0.
$$
This expression is in its turn reducible because, on shell, $$
F_{\eta\nu}\Omega^{\mu\nu\rho} =0,
$$
and the tower of reducibility continues indefinitely. This fact was 
already noticed in \cite{bana96}. Since in our approach we retain all 
the variables (that is, including $A_{0}$), our description is fully 
covariant.

\subsubsection{ Counting the degrees of freedom} 

\noindent{\bf A) In tangent space.}\\
Here we present an alternative counting \cite{bana96} of the degrees 
of freedom for the abelian Chern-Simons theory. Since most of our 
Noether gauge transformations are not projectable, we will study the 
gauge fixing procedure in the Lagrangian formalism as discussed in 
\cite{pons95}. Here we will ignore the fact that the Lagrangian is 
linear in the velocities and we will apply standard machinery. First 
we count the Lagrangian constraints in the usual sense, that is, 
constraints involving configuration and velocity variables. They are 
$K=0, \, {\dot K} =0$ and
$\Omega^{ij}F_{0j} = 0$. This amounts for for a total of $2n+2$ 
Lagrangian constraints. Now we must introduce new constraints in 
order to eliminate the gauge freedom. The independent gauge degrees 
of freedom are described by the arbitrary functions $\epsilon_{0}$ 
and $\eta^{p}$ in (\ref{thegaugetr}); these functions must be fixed 
by introducing the gauge fixing constraints. This means we must 
introduce $2(n-1) + 1$ new constraints to fix the dynamics. At this 
point there is still a redundancy in the setting of the initial 
conditions (residual gauge freedom, see \cite{pons95} for details) 
because, at any given time $t_{0}$, the function $\dot \epsilon_{0}$ 
is an independent function and one can verify there are still gauge 
motions that preserve the gauge fixing constraints. A new gauge 
fixing constraint is necessary. Thus, we have introduced a total of 
$2(n-1) + 2 = 2n$ gauge fixing constraints. The final number of 
constraints is, therefore, $4n+2$. This number equals the dimension 
(in the sense of field theory) of the tangent space. There are no 
local degrees of freedom.

\noindent{\bf B) In configuration space.}\\ Since the Lagrangian 
equations of motion are linear in the velocities, the real setting of 
the initial conditions is in configuration space. So let us do things 
better: We will only consider constraints in configuration space (we 
will call them cs-constraints). The theory only provides with one 
such a constraint, namely $K=0$. 

Using the first relation in (\ref{nice}) (we always consider the 
theory to be generic), the equations of motion, $$
\Omega^{ij}F_{0j} = 0,
$$
can be written as
$$
F_{0\alpha} = F_{0q}F^{qp}F_{q\alpha}.
$$
Thus, $F_{0q}$ and $\dot A_{0}$ are arbitrary. Defining (to coincide 
with the notation used in the Hamiltonian analysis) $$\lambda_{0} := 
\dot A_{0}, \qquad \lambda_{p} := \dot A_{p} - \partial_{p}A_{0} = 
F_{0p},
$$
we can express the evolutionary vector field in a way that encodes 
all the arbitrariness through these functions $\lambda_{0},\, 
\lambda_{p}$: $$
{\bf X_{L}} = {\partial \over \partial t} + \lambda_{0} {\delta \over 
\delta A_{0}} +(\lambda_{p} +\partial_{p}A_{0}){\delta \over \delta 
A_{p}} +(\partial_{\alpha}A_{0} + \lambda_{p}F^{pq}F_{q\alpha}) 
{\delta \over \delta A_{\alpha}}.
$$
As usual, the gauge fixing procedure has two steps: fixing the 
arbitrariness in the dynamics and fixing the redundancy in the 
initial conditions. To fix the dynamics we need to introduce as many 
gauge fixing cs-constraints as arbitrary functions in $X_{L}$, that 
is, $2(n-1) + 1$ cs-constraints. Once the dynamics is fixed we must 
consider the residual gauge freedom that is still available in the 
setting of the initial conditions. The same argument used before 
makes it necessary an additional gauge fixing cs-constraint 
\cite{pons95}. Now the gauge freedom has been completely eliminated. 
We end up with $2n$ gauge fixing cs-constraints to be added to the 
original cs-constraint $K=0$. The final number of cs-constraints, 
$2n+1$, equals the dimension of configuration space.
Again, there are no local degrees of freedom. 



\subsection{Example 5: Pure electromagnetism} 

In this example we consider the construction of the gauge symmetry in 
pure electromagnetism. From the
Lagrangian $${\cal L} = -\frac{1}{4} F_{\mu \nu}F^{\mu \nu}$$ we get 
a canonical Hamiltonian,
$$H_c = \int d{\bf x} \left[\frac{1}{2}({\bf \pi}^2 + {\bf B}^2) + 
{\bf \pi} \cdot \nabla A_0\right],$$ with $ B^k = -\frac{1}{2} 
\epsilon^{0kij}F_{ij}$, and a primary constraint (coming from the 
definition of the momenta) $\pi_0 = 0$. Stability of this constraint 
under the Hamiltonian dynamics leads to the secondary constraint 
$\dot\pi_0 = \{\pi_0, H_c\} = \nabla \cdot {\bf \pi} = 0$, and no 
more constraints arise. Both constraints are first class, so we can 
directly apply the existence theorem of subsection 6.3 to claim that
$$
G = \int d{\bf x} \ \epsilon({\bf x}, t)\nabla \cdot {\bf \pi} $$
is a Noether conserved quantity satisfying (\ref{cond1}) and 
(\ref{cond2}), with $\epsilon$ an arbitrary function. Coefficients in 
(\ref{cond2i}) and (\ref{cond1i}) are, in this case, $A=\dot\epsilon$ 
,
$B=0$. The transformation (\ref{delq}) becomes $$\delta A_\mu= 
-\partial_\mu \epsilon,
$$
which is canonically generated by
$$
G'= G - A \pi_0 = - \int d{\bf x} \ \pi^\mu \partial_\mu \epsilon. $$
This gauge generator, $G'$, satisfies, by construction, 
(\ref{cond2k}) and (\ref{cond1k}).



\section{Conclusions}

In this paper we give results concerning the formulation of Noether 
symmetries, either in the tangent space or in the phase space for 
some configuration space of a dynamical system that may contain gauge 
freedom.

The conserved quantities
associated with the Noether symmetries are characterized in the phase 
space,
and we show the role played by
the Dirac bracket structure for this characterization. We also give a 
geometric property that must be satisfied by a Noether conserved 
quantity in the tangent space in order that the associated Noether 
transformation be projectable to phase space. In such case, the 
Noether transformation becomes a canonical transformation. 

We introduce the enlarged formalism (that includes the Lagrange 
multipliers as independent variables) as a wider framework to deal 
with generalized Noether symmetries. A general formula is obtained 
that fully characterizes the conserved quantities associated to these 
symmetries. We also show how to
bring these Noether symmetries in the enlarged formalism to the 
original Lagrangian.

The algebra for the set of Noether transformations is also discussed,
and it is pointed out that the closure of the algebra of generators 
under the Poisson bracket in phase space does not guarantee the 
closure of the commutator of transformations in configuration space. 
This means that sometimes the presence of an open algebra structure 
will be just an artifact of the pullback from the phase space 
description to the description in configuration space.

Noether transformations in general can be rigid or gauge. We discuss 
some aspects of both types of symmetries and, in the gauge case, we 
give a general proof of the existence of the right number of 
independent Noether gauge transformations for theories with first and 
second class constraints with a stabilization algorithm that does not 
exhibit tertiary constraints. It is still an open problem to prove 
the existence of the right number of independent Noether gauge 
transformations for general theories with an undetermined number of 
steps in the stabilization algorithm. 

The generality of the enlarged formalism suggests us that it is the 
most general framework to deal with Noether symmetries, that is, that 
any Noether symmetry can be cast into the general formula 
(\ref{enl-cond}). Work is in progress to prove this assertion 
\cite{garc-pon99}.

We present some examples to illustrate the application of our 
results. Of particular interest are examples 2 and 4. The first 
because it exhibits a generator of Noether gauge transformations that 
must include terms that are quadratic in the constraints, which is 
uncommon. The second because it is an example of a field theory that 
has a relevant role in the modern developments in quantum field 
theory.

We have worked under standard regularity assumptions: the constancy 
of the rank of the Hessian matrix and the existence of a splitting of 
the primary constraints into first and second class on the primary 
constraint surface. Under these assumptions, here we list our main 
results as presented in the text (except for result 5, that only 
needs the first assumption ):

\begin{enumerate}

\item
{\it The necessary and sufficient condition for a function $G \in 
T^*Q\times R$ to be a Noether canonical conserved quantity, that is, 
such that its pullback $G_{L}$ to $TQ \times R$ satisfies
$$
[L]_i\delta q^i + \frac{dG_L}{dt} = 0,
$$
for some
$\delta q$, is that $G$ must satisfy
\begin{eqnarray*}
\fbox{
$ \displaystyle
\{G,\phi_{\mu_0} \} = sc + pc, \quad
{\partial G\over\partial t} + \{G,H_c\}^* = sc + pc $}
\end{eqnarray*}
where $pc$ ($sc$) stands generically for primary (secondary) 
constraints.} \\ (Section 3, subsection 3.1)

\item
{\it The Noether transformation is reconstructed from $G$ through 
\begin{eqnarray*}
\fbox{
$ \displaystyle
\delta q^{i} = {\cal F}\!L^*\{q^{i},\,G \}^* - ({\cal F}\!L^* 
A^{\nu_0} + v^{\mu_0} {\cal F}\!L^* B^{\nu_0}_{\mu_0}) 
\gamma^{i}_{\nu_0} $}
\end{eqnarray*}
where $B^{\nu_0}_{\mu_0}$ and $A^{\nu_0}$ are the coefficients of the 
$sc$ terms in the preceding expression.}\\ (Section 3, subsection 3.2)

\item

{\it A Noether transformation in phase space is always a canonical 
transformation}.\\
(Section 3, subsection 3.2)

\item

{\it The necessary and sufficient condition for a Lagrangian Noether 
conserved function $G_L$ to be associated through $$
[L]_i\delta q^i + \frac{dG_L}{dt} = 0.
$$
with a transformation $\delta q^{i}$ that is projectable to phase 
space is that $G_L$ must give zero when acted upon by the vector 
fields in the kernel of the presymplectic form in tangent space.} \\
(Section 3, subsection 3.3)

\item

{\it The necessary and suficient condition for a function 
$G^c(q,p,\lambda, \dot \lambda, \ddot \lambda,...;t)$ to generate 
through
$$
\delta_c q ^{i} = \{q^{i}, \, G^c \}, \quad \delta_c p_{i} = \{p_{i}, 
\, G^c \},
$$
a Noether symmetry in the enlarged
formalism is that $G^c$ must satisfy
\begin{eqnarray*}
\fbox{
$ \displaystyle
\frac{D G^c}{D t} + \{ G^c,\, H_D \} = pc $ }
\end{eqnarray*}
where $pc$ stands generically for primary constraints and $$
\frac{D }{D t} :=
\frac{\partial }{\partial t} + {\dot \lambda^\mu} \frac{\partial} 
{\partial \lambda} + {\ddot \lambda^\mu} \frac{\partial} {\partial 
{\dot \lambda}} +... \ ,
$$
}\\
(Section 4)

\item

{\it A sufficient
condition for the commutation relations of the projectable Noether 
transformations in configuration space to coincide with the Poisson 
bracket relations of their generators in phase space is that at least 
all but one of the generators are linear in the momenta}. \\ (Section 
5)

\item

{\it The existence of a rigid Noether conserved quantity does not 
guarantee that its associated Noether transformation is projectable 
to the canonical formalism}. \\
(Section 6, subsection 6.1)

\item

{\it Any theory with
only primary and secondary constraints (all effective) exhibits a 
basis of independent canonical Noether gauge generators in a number 
that equals the number of final primary first class constraints.}\\ 
(Section 6, subsection 6.3)

\end{enumerate}

\section*{Acknowledgments}
We are grateful to Lawrence C. Shepley for useful conversations and a 
critical reading of the manuscript. JAG acknowledges partial support 
from the grants DGPA-IN100397 and CONACyT No. 3141P-E9608 and 
Departament d'Estructura i Constituents de la Mat\`eria, Universitat 
de Barcelona, where part of this work was done. JMP acknowledges
support by CICYT, AEN98-0431, and CIRIT, GC 1998SGR, and wishes to 
thank the Comissionat per a Universitats i Recerca de la Generalitat 
de Catalunya for a grant. He also thanks the Center for Relativity of 
The University of Texas at Austin for its hospitality. 


\appendix

\section*{Auxiliary variables}

Consider a configuration space locally described by the coordinates 
$x_a, y_j$. Suppose a general Lagrangian of the form $$
L(x_a,y_j;\dot x_a,\dot y_j),
$$
and suppose that the $y$ variables are auxiliary variables, that is, 
their equations of motion allow
for the isolation of these variables in terms of $x$ and $\dot x$, $$
[L]_y = 0 \Longleftrightarrow
y = {\hat y}(x, \dot x).
$$
(The components of the two dimensional metric in the Polyakov string 
\cite{polya} are examples of such type of variables) Then define the 
reduced Lagrangian $L_r$ as $$ L_r(x, \dot x, \ddot x) = L{}_{|_{y = 
{\hat y}(x, \dot x)}}. $$
We will prove that $L_r$ gives the right equations of motion for the 
remaining variables $x$.

There is the following relationship between the equations of motion 
for $L_r$ and those for $L$,
$$
[L_r]_x = [L]_x{}_{|_{y = {\hat y}(x, \dot x)}} + \frac{\partial \hat 
y }{\partial x} [L]_y{}_{|_{y = {\hat y}(x, \dot x)}} -\frac{d}{dt} 
\left( \frac{\partial \hat y}{\partial \dot x} [L]_y{}_{|_{y = {\hat 
y}(x, \dot x)}} \right),
$$
but since
$$
[L]_y{}_{|_{y = {\hat y}(x, \dot x)}} =0 $$
identically, we conclude that
$$
[L_r]_x = [L]_x{}_{|_{y = {\hat y}(x, \dot x)}}. $$
This proves that the dynamics for the reduced variables $x$ is 
governed by the reduced Lagrangian $L_r$. 

When we apply this result to section 4, the Lagrangian $L_c$ from 
which we start the reduction procedure does not depend on the 
velocities $\dot y$. The $y$ variables are, in the notations of 
section 4, $p$ and $\lambda$.



\end{document}